\documentclass[usenatbib,a4paper]{mn2e}
\usepackage[totalwidth=510pt, totalheight=680pt]{geometry}
\usepackage{times}
\usepackage{graphicx}
\usepackage{graphics}
\usepackage{rotating}
\usepackage{amsmath}
\usepackage{amssymb}
\title[Radio sources in the 6dFGS]
  {Radio sources in the 6dFGS: Local luminosity functions at 1.4\,GHz
    for star-forming galaxies and radio-loud AGN.}

\author[T. Mauch and E. M. Sadler]
{Tom~Mauch$^1$\thanks{Present address: Astrophysics, Department of Physics, 
    Denys Wilkinson Building,
    University of Oxford, Keble Road, Oxford OX1 3RH, UK. 
   E-mail: txm@astro.ox.ac.uk}
 and Elaine~M.~Sadler$^1$\\
$^1$School of Physics,
  University of Sydney,
  NSW, 2006, Australia.\\
}

\date{Accepted 2006 November 27. Received 2006 November 23; in original form 2006 October 4}

\pagerange{\pageref{firstpage}--\pageref{lastpage}}

\pubyear{2006}

\begin{document}

\maketitle

\label{firstpage}

\begin{abstract}
We have identified 
7824 radio sources
from the 1.4\,GHz NRAO VLA Sky Survey (NVSS) with 
galaxies brighter than $K=12.75$\,mag. in the Second Incremental Data Release
of the 6dF Galaxy Survey (6dFGS\,DR2). The resulting sample
of redshifts and optical spectra for radio sources over an 
effective sky area of 7076\,deg$^{2}$ (about 17\,per\,cent of the celestial
sphere) is the largest of its kind
ever obtained. NVSS radio sources associated with galaxies
in the 6dFGS span a redshift range
$0.003<z<0.3$ and have median $\tilde{z}=0.043$.
Through visual examination of 6dF spectra we have
identified the dominant mechanism for
radio emission from each galaxy. 60\,per\,cent are fuelled
by star-formation and 40\,per\,cent are fuelled by
an active galactic nucleus powered by a supermassive black
hole.
We have accurately determined the local radio luminosity
function at 1.4\,GHz for both classes of radio source and have found
it to agree well with other recent determinations. From the radio
luminosity function of star-forming galaxies we derive
a local star formation density of 
$0.022 \pm 0.001$\,M$_\odot$\,yr$^{-1}$\,Mpc$^{-3}$, in broad agreement
with recent determinations at radio and other wavelengths. 

We have split the radio luminosity function of radio-loud
AGN into bins of absolute $K$-band magnitude
($M_K$) and compared this with the underlying $K$-band
galaxy luminosity function of all 6dFGS galaxies to determine the bivariate
radio-$K$-band luminosity function. 
We verify that radio-loud AGN preferentially inhabit the brightest and hence 
most
massive host galaxies and show that the fraction of all galaxies which host
a radio-loud AGN scales as $f_{\rm radio-loud} \propto L_K^{2.1}$ for $f_{\rm radio-loud}<0.3$,
indicative of a similarly strong scaling with black hole mass and stellar mass.

\end{abstract}


\section{Introduction}

In recent years a new generation of radio surveys has been released
covering a large
fraction of the celestial sphere down to 
flux density levels of a few mJy
(eg. NRAO VLA Sky Survey (NVSS), \citet{nvss}; Sydney University Molonglo Sky Survey (SUMSS), 
\citet{bock99} \& \citet{sumss}; Faint Images of the Radio Sky at Twenty-cm (FIRST), \citet{first};
Westerbork Northern Sky Survey (WENSS), \citet{wenss}). The vast majority of the radio sources in these
surveys are produced by
Active Galactic Nuclei (AGN) powered by supermassive black holes in galaxies
with a median redshift of $z\approx0.8$ \citep{nvss}. These
dominate the radio source 
population above flux densities of 10\,mJy. Below 10\,mJy the
surveys contain an increasing fraction of nearby ($z<0.1$) 
galaxies whose radio emission is fuelled by ongoing star formation \citep{condon89}. 
Because
radio surveys probe a wide range of redshifts, studying their global
properties statistically (eg. through radio source counts or luminosity
functions) provides a
powerful constraint on the evolutionary properties of massive galaxies
throughout the history of the
universe. Unfortunately, radio survey data alone are not sufficient to
constrain the current models of radio source counts. These models are
strongly dependent on the form of the \textit{local} 
radio luminosity function (RLF)
from which the measured source counts can be
extrapolated via evolutionary models \citep[eg.][]{dunlop90,jackson99} or compared
directly to the
measured radio luminosity
function of samples selected at higher redshift \citep[eg.][]{brown01,sadler06}. 
The current generation of redshift
surveys (eg. Sloan Digital Sky Survey (SDSS), \citet{sloan}; 2 degree Field Galaxy Redshift Survey (2dFGRS), 
\citet{2df}; 6 degree Field Galaxy Survey (6dFGS), \citet{jones04}) 
provide a useful tool for calculating the local radio luminosity
function, as they provide redshifts for thousands of radio
sources in the local universe. The host galaxy 
spectra provide additional value in this context
as they can be used to determine the physical cause of the radio emission
from galaxies, thereby disentangling the
star-forming galaxy population from the radio-loud AGN.

Table~\ref{samplecompare} compares the underlying radio-optical 
samples used in recent measurements of the local radio luminosity function.
\citet{2dfnvss} identified NVSS radio sources in the 2dFGRS
and derived the local radio luminosity function of
both star-forming galaxies and radio-loud AGN; this will hereafter be referred to as
the 2dFGRS-NVSS sample. \citet{maglio02} identified optical
counterparts to FIRST galaxies in the 2dFGRS; this will hereafter be referred to
as the 2dFGRS-FIRST sample. Both the radio-selected 2dFGRS samples cover a relatively
small area of sky to redshifts of $z\approx0.3$ and predominantly
comprise radio-loud AGN. In particular the 2dFGRS-FIRST sample detected relatively
few star-forming galaxies as the high-resolution 
FIRST survey resolves out much of the extended radio emission 
in the disks of nearby galaxies. \citet{best05} identified NVSS and FIRST 
radio sources in the second data release of the SDSS; this will hereafter be referred 
to as the SDSS-NVSS/FIRST sample. They optimised
their radio-source selection via a hybrid method which used 
both the resolution of the FIRST survey 
and the surface brightness sensitivity of the NVSS survey. 
However, they detected relatively few
star-forming galaxies above their adopted flux density limit of 5\,mJy.
\citet{condon02} crossmatched the 
NVSS catalogue with galaxies brighter than 
$m_p=14.5$\,mag. in the Uppsala Galaxy Catalogue 
\citep[UGC;][]{ugc} and constructed a
a sample of 1966 radio sources in the local $(z<0.03)$ universe across
4.33 sr of the northern sky; this will hereafter be referred to as
the UGC-NVSS sample. The wide yet shallow UGC-NVSS sample contained a
large population of star-forming galaxies but relatively
few radio-loud AGN. Larger numbers of radio-loud AGN are found at
higher redshifts. 

This paper presents the local radio luminosity function derived 
from a well-defined subsample of the 7824 radio sources from the NVSS
catalogue identified with galaxies observed in the Second Incremental Data Release (DR2) 
of the 6dFGS; this will 
be referred to as the 6dFGS-NVSS sample. Galaxies in the 6dFGS-NVSS sample presented in this paper
lie at redshifts intermediate between the more distant 2dFGRS-NVSS and 
SDSS-NVSS/FIRST samples and
the nearby UGC-NVSS sample, in an effective sky area much larger than that 
of the 2dFGRS and SDSS
derived samples. 6dFGS-NVSS galaxies are contained in 
a larger volume of space than any previous local radio source
sample, with significant populations of both radio-loud AGN and
star-forming galaxies all contained in the one survey.

\begin{table*}
\begin{center}
\caption{Comparison of recent radio-selected samples used to derive the local RLF.}
\label{samplecompare}
\begin{tabular}{lrccccc}
\hline
 & & UGC-NVSS & 6dFGS-NVSS & 2dFGRS-NVSS & 2dFGRS-FIRST & SDSS-NVSS/FIRST \\
 & & \citet{condon02} & This paper & \citet{2dfnvss} & \citet{maglio02} & \citet{best05} \\
\hline
 & & \multicolumn{5}{c}{\bf Star-forming galaxies} \\
\cline{3-7}
 & No. Galaxies & 1672 & 4006 & 242 & 177 & 497 \\
 & Mag. limit & $m_p<14.5$ & $K_{\rm tot}<12.75$ & $14.0 \leq b_J \leq 19.4$ 
& $14.0 \leq b_J \leq 19.45$ & $ 14.5 < r < 17.77$ \\
 & $S_{\rm lim}$ (mJy) & 2.5 & 2.8 & 2.8 & 1.0 & 5.0 \\
 & Area(Sr.)$^{\rm a}$ & 4.33 & 2.16 & 0.10 & 0.07 & $\sim0.4^{\rm e}$ \\
 & $z$ range$^{\rm b}$ & $\lesssim 0.04$ & $\lesssim 0.1$ & $\lesssim 0.15$ & $\lesssim 0.25$ & $\lesssim 0.15$ \\
 & median $z$ & 0.012 & 0.035 & 0.043 & 0.100 & 0.055 \\
 & Volume ($10^{6}$Mpc$^3$)$^{\rm c}$ & 7 & 53 & 8 & 22 & $\sim30^{\rm e}$ \\
 & $\tau$\,(Gyr)$^{\rm d}$ &  0.543 & 1.30 & 1.89 & 2.94 & 1.89 \\
 & $\rho_{\rm SF}$\,M$_{\odot}$\,yr$^{-1}$\,Mpc$^{-3}$ & 0.018 & 0.022 & 0.031 & & \\
\hline
 & & \multicolumn{5}{c}{\bf Radio-loud AGN} \\
\cline{3-7}
 & No. Galaxies & 294 & 2661 & 420 & 372 & 2215 \\
 & Mag. limit & $m_p<14.5$ & $K_{\rm tot}<12.75$ & $14.0 \leq b_J \leq 19.4$ &  $14.0\leq b_J \leq 19.45$ & $ 14.5 < r < 17.77$ \\
 & $S_{\rm lim}$ (mJy) & 2.5 & 2.8 & 2.8 & 1.0 & 5.0 \\
 & Area(sr.)$^{\rm a}$ &  4.33 & 2.16 & 0.10 & 0.07 & $\sim0.4^{\rm e}$ \\
 & $z$ range$^{\rm b}$ & $\lesssim 0.04$ & $\lesssim 0.2$ & $\lesssim 0.3$ & $\lesssim 0.3$ & $\lesssim 0.3$ \\
 & median $z$ & 0.019 & 0.073 & 0.140 & 0.150 & 0.165 \\
 & Volume ($10^6$Mpc$^3$)$^{\rm c}$ & 7 & 391 & 58 & 40 & $\sim 230^{\rm e}$ \\
 & $\tau$\,(Gyr)$^{\rm d}$ & 0.543 & 2.44 & 3.44 & 3.44 & 3.44 \\
\hline
\end{tabular}
\end{center}
\begin{flushleft}
\footnotesize
NOTES: \\
$^{\rm a}$ For the 2dFGRS-NVSS and 6dFGS-NVSS samples the area is the
effective area as defined in Section~\ref{primcov}.\\
$^{\rm b}$ The maximum redshift of each survey has been estimated from the
redshift histogram.\\
$^{\rm c}$ Volumes are calculated to the maximum redshift of each survey and
reduced by the fraction of sky surveyed.\\
$^{\rm d}$ $\tau$ is the lookback time in Gyr and is calculated from the maximum
redshift of each survey.\\
$^{\rm e}$ As \citet{best05} did not calculate the sky area of the SDSS-NVSS/FIRST sample, 
we have roughly estimated it from coverage plots.\\
\end{flushleft}
\end{table*} 

The radio luminosity function derived in this paper
has a number of advantages over previous determinations:
\begin{itemize}
\item The Near-Infrared (NIR) and radio input
catalogues are from homogeneous surveys
with a single instrument, and avoids
biases which may result from combining surveys. 

\item The digital 
$K$-band magnitudes from the 2MASS\,XSC are more accurate than those which
have been derived from measurements of photographic plates (eg. 2dFGRS, UGC). 

\item The near-infrared selection of the 6dFGS means it is 
relatively unaffected by dust in both the target galaxy and our own Galaxy. 
This means that the radio selected
sample will not be biased with respect to the amount of dust in the host
galaxy, which can steer surveys away from galaxies with higher
star-formation rates. Near-infrared magnitudes are also more closely linked to 
the old stellar population in galaxies, effectively tipping the balance 
in favour of massive early type galaxies which preferentially
host radio-loud AGN.

\item The large sample volume of the 6dFGS contains more radio sources than
any previous radio-selected galaxy sample (see table~\ref{samplecompare}).
\end{itemize}

The structure of this paper is as follows. Section~\ref{sampleselec} describes the 6dFGS and
NVSS surveys and our method of identifying radio sources in the 6dFGS. 
Section~\ref{primsamplesec}
outlines the global properties of 6dFGS-NVSS
galaxies. Section~\ref{primiras} examines the population of primary sample
objects which are detected in the IRAS-FSC, both as a consistency check for
the spectroscopic classification of the sample and to derive the radio-FIR
correlation for a larger sample of star-forming galaxies than any
obtained to date. 
Section~\ref{primlumfunc} presents the local radio luminosity
function at 1.4\,GHz for the radio-loud AGN and star-forming galaxies.
Section 6 presents a derivation
of the star-formation density at the present epoch. 
Section~\ref{fraclumfunc} presents the bivariate radio-NIR luminosity
function of radio-loud AGN. Finally,
Section~\ref{primconclusion} summarises the main results. 
Throughout, if not explicitly stated, we adopt a $\Lambda$CDM cosmology
with parameters $\Omega_{\rm m}=0.3$,
$\Omega_{\rm \Lambda}=0.7$, and $H_0=70$\,km\,s$^{-1}$\,Mpc$^{-1}$ \citep{spergel03}.

\section{Sample Selection}
\label{sampleselec}

\subsection{The 6 degree Field Galaxy Survey}

The 6 degree Field Galaxy Survey
is a spectroscopic survey of about 170\,000 objects in 
the 17\,046\,deg$^2$ of sky with declination $\delta < 0^\circ$ and 
galactic latitude $|b| > 10^\circ$. 
The survey was carried out on the 1.2\,metre UK\,Schmidt Telescope at
Siding Spring Observatory. 
The majority of targets (113\,988) are near-infrared selected to be complete
to an apparent magnitude limit of 
$K=12.75$\,mag. from the Two Micron All Sky Survey Extended 
Source Catalogue \citep[2MASS\,XSC;][]{2mass} and comprise the 6dFGS \textit{primary sample}. 
The remainder of the survey is filled out by a number of smaller
additional target samples selected in various ways at other wavelengths \citep{jones04}. 
The spectroscopic sample considered in this paper consists of
the 47\,317 objects from the $K$-selected primary sample that have been
observed during the period  2002\,January to 
2004\,October and released in the Second Incremental Data Release \citep[DR2;][]{6dfdr2} 
of the 6dFGS and that overlap with
southern part of the NVSS survey ($-40^\circ<\delta<0^\circ$). 

6dFGS spectra are obtained through 6\,arcsec diameter 
fibre buttons which correspond to 
a projected diameter of 6.8\,kpc at the median redshift of the survey 
($\tilde{z}\sim0.05$).
6dF fibres include an increasing fraction of total galaxy light for higher
redshift galaxies, and therefore galaxies with emission-line nuclei are
easier to recognise at lower redshift.
6dFGS observations prior to 2002\,October were made with reflection gratings
and cover a wavelength range of 4000--8400\AA. Since 2002\,October
observations were made with Volume-Phase transmissive Holographic (VPH) gratings, 
which have provided improved efficiency
and data uniformity and cover a wavelength range of 3900--7500\AA. Primary target
spectra typically have
resolution $R\sim1000$ and signal-to-noise ratio (S/N) of 10\,pixel$^{-1}$ allowing
redshifts and host galaxy parameters (ie. emission-line ratios) to be readily
determined. Redshifts are measured from the spectra using the {\sc runz} package
described by \citet{jones04} and assigned a quality class $Q$ by human operators
based on the reliability of the measured redshift. Only $Q=3$ or $4$ redshifts
are expected to be reliable, $Q=6$ redshifts are reserved for Galactic sources
\citep[see][for a complete description]{6dfdr2}.

The total $K$-band magnitudes ($K_{\rm tot}$) used for selection of 
6dFGS primary targets have been derived from isophotal magnitudes ($K_{\rm iso}$)
and diameters to an elliptical isophote of $\mu_K=20$\,mag.\,arcsec$^{-2}$ 
\citep[equation 1 of][]{jones04}.
These $derived$ values of $K_{\rm tot}$ are more robust than those quoted
in the 2MASS\,XSC and have therefore been used throughout this paper. 
The derived $K$ magnitudes have typical errors $\Delta K<0.1$\,mag. and the near-infrared-selected
sample is complete to the survey limit of $K=12.75$\,mag.~\citep{jones06}.

\subsection{The NRAO VLA Sky Survey}

The NVSS is a radio imaging survey of the entire sky north of
$\delta=-40^\circ$ at 1.4\,GHz carried out on the Very Large Array (VLA) telescope in its DnC
and D configurations. 
Its principal data products are a set of 2326 $4^\circ \times 4^\circ$ continuum 
images and a catalogue of about $2 \times 10^6$ fits of elliptical gaussians to
discrete sources stronger than $S_{1.4\,{\rm GHz}}\approx2\,{\rm mJy}$. 
The images have $\theta = 45\,{\rm arcsec}$ FWHM
resolution with position accuracy $\leq 1''$
for sources stronger than 15\,mJy, 
increasing to $>7''$ at the survey limit \citep{nvss}. The
resolution and positional accuracy of the NVSS
makes identification of radio sources with objects on optical
survey plates straightforward in the majority of cases. 
There are 580\,419 NVSS catalogue sources
which overlap with the 6dFGS; a region bounded by $0^\circ>\delta>-40^\circ,
|b|>10^\circ$. These sources form the primary 1.4\,GHz-selected
input catalogue for the 6dFGS-NVSS sample.

\subsection{Sky coverage}
\label{primcov}

The 6dFGS uses a tiling algorithm with variable overlap depending
on the underlying galaxy density \citep{campbell04}. Because of this 
and because not all 6dFGS tiles were observed in the DR2 \citep{6dfdr2}, 
calculating the exact area of sky surveyed is not straightforward. A method
of calculating sky area for surveys with gaps in coverage such as
the 6dFGS\,DR2 was described by \citet{folkes99} and was also applied to 
2dFGRS-NVSS galaxies by \citet{2dfnvss}. This method
estimates the area of sky covered by dividing the number of galaxies
observed by the mean surface density of galaxies in the survey target
list. The surface
density of 6dFGS primary targets is 6.7 objects\,deg$^{-2}$. 
47\,317 primary targets north of $-40^\circ$ have been observed in the DR2. 
Therefore the effective
area is 7\,076\,deg$^2$, or 17.15\,per\,cent of the celestial
sphere. Calculation of the effective area in this way also accounts for the
5\,per\,cent 
fibreing incompleteness of the 6dFGS. However, it takes no account of
the small amount of
incompleteness due to failed observations (eg. low S/N spectra); this is
discussed in more detail in Section~\ref{primlumfunc}.

\subsection{Crossmatching the 6dFGS primary sample with the NVSS}

When searching for NVSS radio source identifications of 
the 6dFGS primary targets
we used a method which aimed to maximise both the completeness
(ie. all potential radio sources are included) and the reliability (ie. all
included radio identifications are genuine) of the database.
To ensure a high completeness and reliability of NVSS-2MASS
identifications we
determine a maximum position offset which includes all
possible identifications, construct a list of these identifications and
then verify each of them by visually inspecting overlays of radio contours onto 
optical images.
Although checking all identifications by eye is time consuming, in our experience
it is the best way to minimise the number of false identifications, as at all
separations
some radio identifications selected by position offset alone will be chance
coincidences. Visual inspection of every radio-optical
overlay also allows us to correct
the catalogued flux densities of confused or diffuse sources for which the
elliptical gaussian model used in the NVSS catalogue does not measure
reliable radio 
source parameters. Our visual identification method closely follows that
described by both \citet{condon02} \& \citet{2dfnvss}; the reader is referred to those 
papers for examples and descriptions of radio source identifications with bright galaxies 
in optical images.

To include all possible radio source identifications, we decided to consider all 
radio sources with NVSS-2MASS position offset less than 3\,arcmin.
This offset was chosen as it is large enough to include the majority 
of possible multi-component radio galaxies while still
smaller than the average separation of unrelated radio-sources in the 
NVSS catalogue. Of
the 74\,609 6dFGS primary sample targets north of 
$-40^\circ$, 34\,097 of them 
had at least one identification in the NVSS catalogue within 3\,arcmin, of which 
21\,597 were observed in the 6dFGS\,DR2.
We chose to examine the 15\,716 6dFGS\,DR2 galaxies that have a single NVSS
identification within 3\,arcmin separately from the 5\,881 with more than one.

\subsubsection{Single-component radio sources}
\label{singlecomp}

\begin{figure}
\centering
\includegraphics[width=\linewidth]{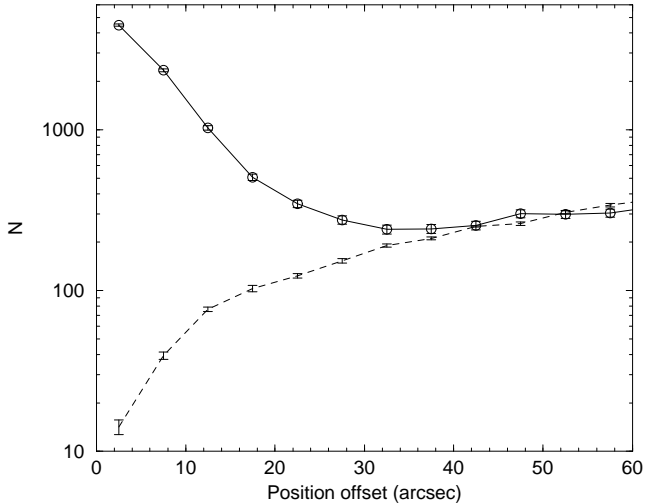}
\caption{For all 25\,096 
  6dFGS primary targets with a single NVSS catalogue source within 3\,arcmin, the open circles linked by a 
  solid line show the distribution of position offsets to the radio source.
  The dashed line shows the average distribution of position offsets to NVSS catalogue sources from positions 
  in 5 random catalogues with the same size and sky coverage as the 6dFGS primary sample, 
  also with a single radio source within 3\,arcmin.
}
\label{listin1}
\end{figure}

\begin{figure}
\centering
\includegraphics[width=\linewidth]{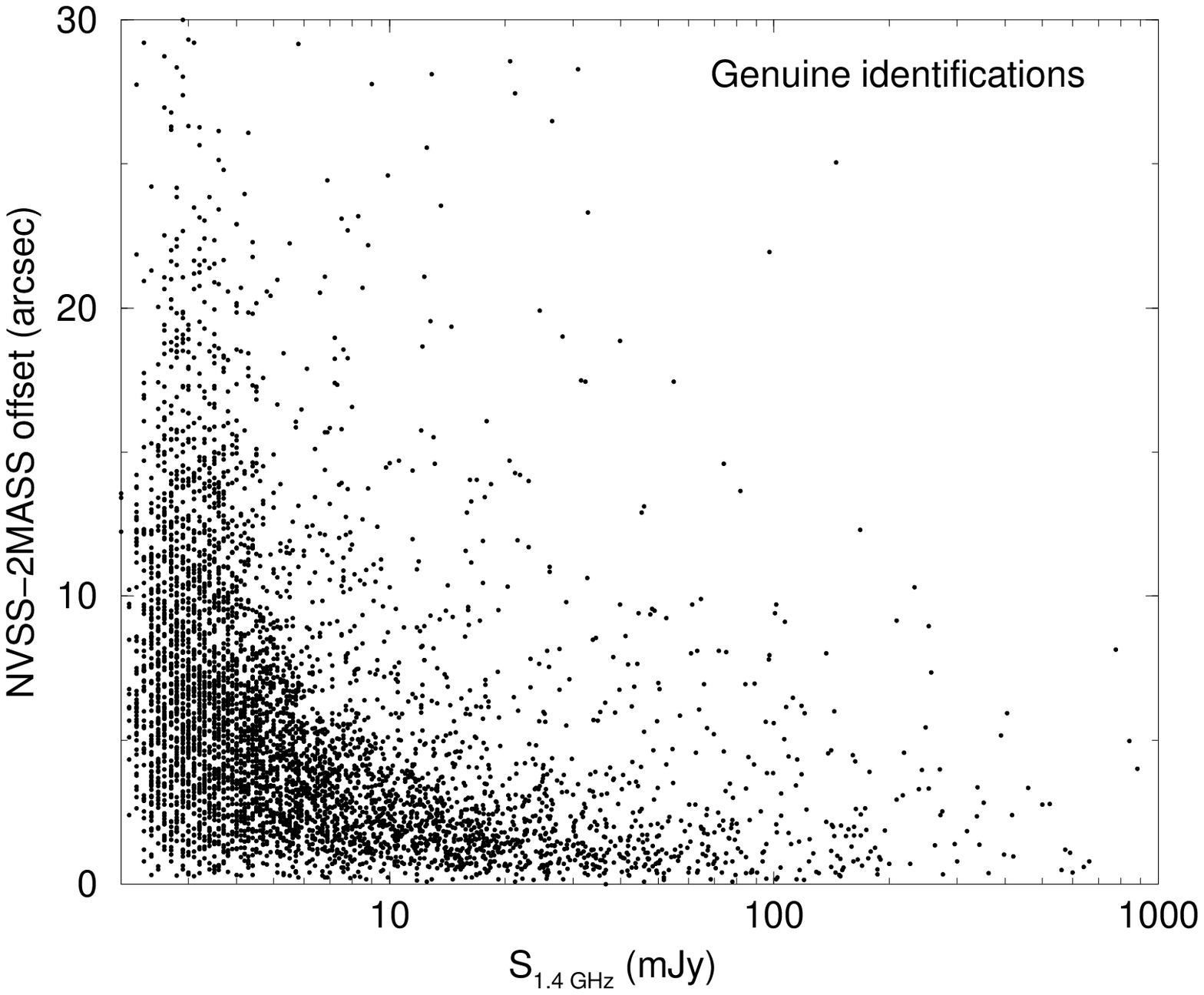}
\includegraphics[width=\linewidth]{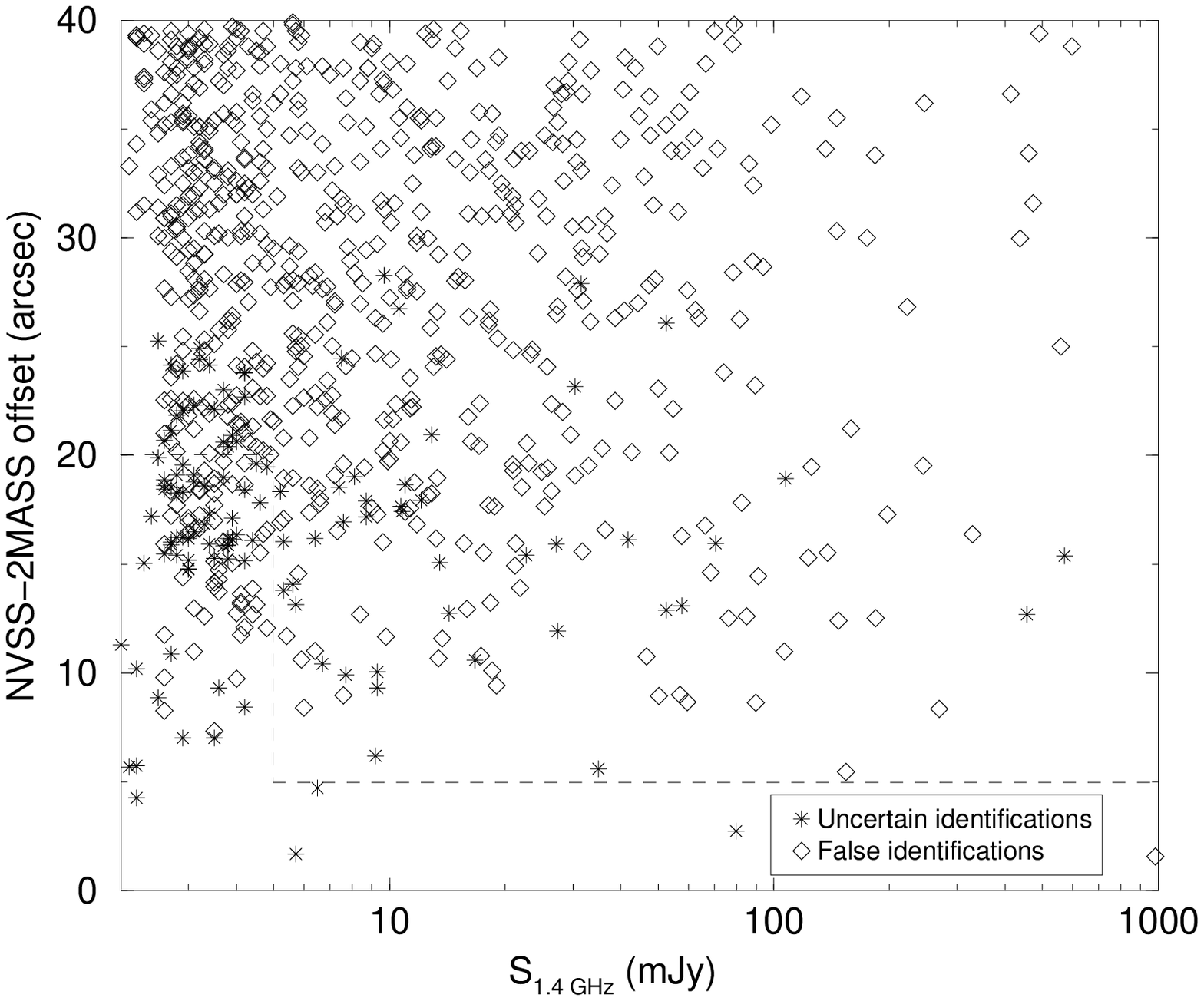}
\caption{The results of visual identification of
  $5\,829$ 6dFGS\,DR2 primary sample objects with a single NVSS catalogue match
  within 3\,arcmin.
  The top figure shows the 5\,016 real
  identifications, we found no genuine identifications with NVSS-2MASS offset $>30$\,arcsec. 
  In the bottom figure, diamonds show
  the 693 false identifications and
  stars show the 120 uncertain identifications. The dashed line shows
  the adopted cutoff used to accept uncertain identifications.}
\label{fluxoffset}
\end{figure}

\begin{figure*}
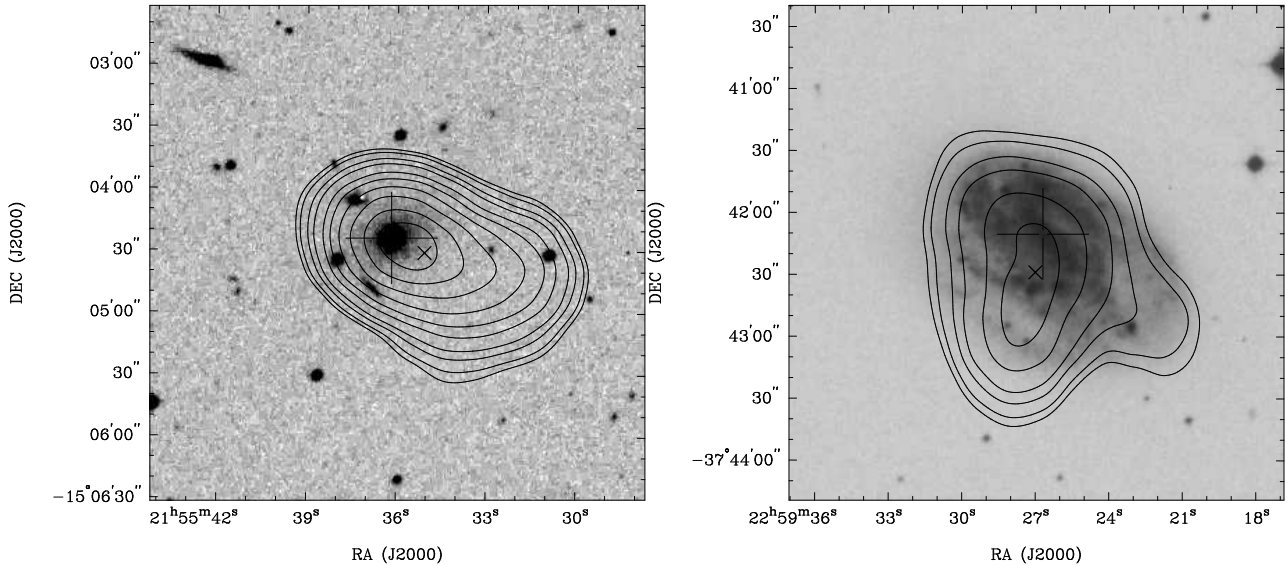

\includegraphics[width=8.5cm]{NJ215535-150431.fits.ps}%
\includegraphics[width=8.5cm]{NJ225927-374229.fits.ps}
\caption{Radio contours from the NVSS overlaid onto $b_J$ images from the 
  SuperCOSMOS Sky Survey for two accepted 6dFGS-NVSS identifications 
  with fitted positions in the NVSS catalogue more than 15\,arcsec from galaxy
  positions in the 2MASS\,XSC. 
  In both images crosshairs show the 
  2MASS\,XSC galaxy position and a small cross labels the fitted position of 
  the radio source in the NVSS catalogue. The contour levels plotted 
  in both images are: 
  2.5, 3, 4, 5, 7, 9, 12, 16, 21, 26 \& 30\,mJy\,beam$^{-1}$.
  The left image shows the 6dFGS galaxy 
  g2155362-150424 at $z=0.07$. The catalogued $S_{1.4\,{\rm GHz}}=55$\,mJy
  NVSS radio source is 17.5\,arcsec from the host galaxy position. Its 
  Aa spectrum (as defined in Table~\ref{claspec}) implies that this is a 
  radio-loud AGN and all of the
  extended radio emission is associated with the host galaxy. The right image
  shows the 6dFGS star-forming galaxy g2259267-374210 at $z=0.004$. The catalogued 
  $S_{1.4\,{\rm GHz}}=39.8$\,mJy NVSS radio source
  is 18.9\,arcsec from the host galaxy position. 
  This source has a complex radio structure and 
  we have therefore corrected its 1.4\,GHz flux density to 42.5\,mJy
  by adding up the pixels in the radio image.}
\label{offsetex}
\end{figure*}

For all 25\,096 6dFGS primary targets with a single NVSS catalogue match within 3\,arcmin, 
the solid line in 
Fig.~\ref{listin1} shows the distribution of position offsets of radio sources
from the 6dFGS galaxy (in arcsec). Five random catalogues containing 74\,609 positions
(the same as the number of all 6dFGS primary targets) and covering 
the same region of sky as the 6dFGS-NVSS sample were also matched with the NVSS catalogue.
For the random positions with only a single NVSS match within 3\,arcmin, 
the average of the 5 resulting distributions of NVSS-2MASS position offsets
is shown as a dashed line in Fig.~\ref{listin1}. At 40\,arcsec the expected
number of
random matches is the same as the number of real matches so we conclude that all
single-component NVSS matches with 6dFGS primary targets that have position offset
$>40$\,arcsec are not genuine.

5\,829 6dFGS\,DR2 primary targets with a single NVSS catalogue match within 3\,arcmin
have position offset $<40$\,arcsec. The dashed line in Fig.~\ref{listin1} implies 
that at separations $<40$\,arcsec some misassociations are expected.
We visually inspected overlays of
NVSS radio contours onto $b_J$ images from the SuperCOSMOS Sky Survey~\citep{supercosmos}
for each candidate identification and classified them as follows
\begin{itemize}
\item \textit{Genuine} (5\,016 candidates), for objects whose radio-optical identification is unambiguously
  real.
\item \textit{False} (693 candidates), for objects whose radio-optical identification is unambiguously not 
  real. This is usually because the radio source is identified with another galaxy on the
  optical image.
\item \textit{Uncertain} (120 candidates), for objects whose radio optical identification is ambiguous. This
is usually because two possible host galaxies can be seen within the NVSS position error ellipse.
\end{itemize}

Fig.~\ref{fluxoffset} shows our hand classification of 6dFGS-NVSS 
candidates. We find no genuine identifications of single component radio
sources with position offset between 30\,arcsec
and 40\,arcsec. The excess of real over random identifications of NVSS radio sources
at these separations seen in Fig.~\ref{listin1} 
is likely to have been caused by the clustering of galaxies
as was noted by \citet{best05}; at a real galaxy position, there is more likely
to be a companion which is the true radio source identification within 100\,arcsec than 
at a random position on the sky.
For accepted NVSS-2MASS identifications
the increasing range of position offsets with decreasing NVSS flux density
reflects the increase in position uncertainties for fainter radio sources \citep{nvss}.
It is clear that there is no
simple NVSS-2MASS offset cutoff which separates real identifications from
false identifications. The vast majority of real identifications have small
(ie. $<15$\,arcsec) position offsets
though there is still about 2\,per\,cent contamination by false
identifications with smaller separations at all NVSS flux densities. Conversely,
some real identifications are found with offset $>15$\,arcsec.
This is due to a number of factors:

\begin{itemize}

  \item Complex and asymmetric radio sources can have fitted positions in the
    NVSS catalogue which differ by more than 15\,arcsec
    from the galaxy positions in the 2MASS\,XSC. Two examples are shown in 
    Fig.~\ref{offsetex}.
    
  \item Some galaxies can appear at the position of one of the lobes of a double
    radio source whose centroid is associated with another galaxy.
    This can lead to false identifications of radio sources with
    offsets $<15$\,arcsec.

  \item One of a pair of galaxies with small separation
    can be identified with a radio source which is clearly identified with
    the other galaxy. This can lead to false identifications of radio
    sources with offsets $<15$\,arcsec.

\end{itemize}

We found 120 candidate identifications which had two or more optical galaxies visible 
on SuperCOSMOS $b_J$ images inside the $3\sigma$ NVSS position error ellipse
(shown as stars in Fig.~\ref{fluxoffset}).
These candidates were considered too uncertain to make a definitive 
visual classification.
We chose to accept as a genuine match the 55 objects with position offset 
$<20$\,arcsec for radio sources with $S_{1.4\,{\rm GHz}} \leq 5\,{\rm mJy}$
and position offset $<5$\,arcsec for radio sources
with $S_{1.4\,{\rm GHz}} > 5\,{\rm mJy}$. This cutoff is shown as a dashed line
in Fig.~\ref{fluxoffset}. In the region of the figure below this line
more than 99 per cent of unambiguous identifications (ie. both genuine and false)
are genuine, so we expect
that less than 1\,per\,cent of unsure identifications will be 
misidentified by applying the cutoff.

\subsubsection{Multiple-component radio sources}

A further 5\,881 6dFGS\,DR2 galaxies had more than one identification in the NVSS
catalogue within 3\,arcmin. We chose to examine all such candidates by visual
examination of SuperCOSMOS $b_J$ images overlaid with NVSS contours. 
In 3\,226 of them, one of the radio components lay within 40\,arcsec of the galaxy. These
were either the cores of multiple-component radio sources or a single component radio
source and other unrelated NVSS radio sources within 3\,arcmin.
We chose to accept or reject such candidates using the method
outlined in Section~\ref{singlecomp} for single component radio sources and
accepted further 2\,716 radio source identifications. We
determined NVSS flux densities for accepted multiple-component radio sources by summing
the individual catalogued flux densities of each component.

We visually classified the remaining 2\,655 candidate multiple-component radio sources which
have no NVSS catalogue match within 40\,arcsec of the host galaxy.
Classification of such objects by eye is straightforward as the sensitivity
of the NVSS to extended radio emission makes it easy to identify large connected 
structures in the radio images.
For brighter radio sources, we supplemented our visual classification by double-checking
our identifications with previous optical identifications of radio sources from the 
Revised Third Cambridge catalogue of radio sources~\citep[3CR;][]{LRL}, 
the Parkes-MIT-NRAO survey catalogue~\citep[PMN;][]{pmn} and the 
Molonglo Reference Catalogue of radio sources~\citep[MRC;][]{mrc81}. A further 37 multiple-component
radio sources were identified out of the 2\,655 candidates.

\subsubsection{Giant radio galaxies}
\label{GRGs}

A detection of
at least one NVSS radio component within 3\,arcmin of the host galaxy
is required to identify a radio source using our identification procedure.
This implies that we miss any population of Giant Radio Galaxies (GRGs)~\citep{sc01,boyce05}
which have linear sizes in excess of 6\,arcmin in radio images and no detection of a core.
About half of the GRGs in a sample selected by \citet{boyce05}
have lobes that are detected at the limit of the SUMSS catalogue but cores that are not. 
The space
density of GRGs with linear size greater than 5\,arcmin measured by \citet{boyce05}
is $10^{-7}$\,Mpc$^{-3}$ for $z<0.4$, which implies that about 40 such GRGs would be
found in the 6dFGS-NVSS sample volume. About 1/3 of these would be missed by
our identification procedure (ie. those with linear size $>6$\,arcmin and 
no detection of a core). GRGs are powerful radio sources and have 
typical 1.4\,GHz radio powers between $10^{23.5}$ and $10^{26}$\,W\,Hz$^{-1}$. We expect 
that
the incompleteness of the 6dFGS-NVSS sample in this radio power range will be
no greater than 1\,per\,cent because of missing GRGs.

\subsubsection{Reliability and completeness}

It is possible to estimate the reliability and completeness 
of the database for single component radio sources
using the data presented in Fig.~\ref{listin1}. 
We expect our visual classification
scheme to be most robust for identifications of radio sources within 15\,arcsec of the host galaxy 
position.
Integrating under the dashed line in Fig.~\ref{listin1} out to 15\,arcsec
and scaling to the number of objects observed implies that 81 random associations with
radio sources are expected. During visual classification we rejected 74 candidate matches
and therefore an estimated 7 accepted identifications in this region are spurious.
The remainder of the accepted matches had offsets between 15\,arcsec and 30\,arcsec.
We expect 238 random associations with offsets in this region and during visual identification
rejected 361 candidate matches. This implies that 123 genuine
identifications in this region 
have been spuriously classified as false. We believe that the majority of these
spuriously rejected identifications have been made for fainter ($S_{1.4\,{\rm GHz}}<2.8$\,mJy) radio sources
many of whose NVSS position errors exceed 10\,arcsec.
Finally, 
we accepted no radio source identifications with position offset between 30 \& 40 arcsec, 
however based on
the small excess of real over random matches seen in this region of Fig.~\ref{listin1}
we expect there to be 51 genuine associations. This means that for 
single-component radio sources in the final database,
7/5016 accepted matches are spurious, corresponding to a reliability of close to 
100\,per\,cent, and
(123+51)=174 matches have been erroneously rejected, corresponding to a
completeness of 97\,per\,cent. Our estimate of completeness is likely to be a lower
limit because many of the excess real over random matches at separations greater than 15\,arcsec are 
associated with galaxy pairs as was noted in section~\ref{singlecomp}. The reliability and completeness of the
2\,716 identifications with more than one NVSS catalogue identification within 3\,arcmin and a radio component
within 40\,arcsec of the host galaxy mimics that of the single-component radio sources.

The reliability and completeness of
multiple-component radio source identifications with no NVSS detection with 40\,arcsec of the host galaxy
is difficult to determine. For the brightest sources, 
we crossmatched the 6dFGS-NVSS database 
with well-studied databases of optical identifications from the 3CR, PMN and MRC databases 
and found no spurious or missing identifications which implies our completeness and reliability
are 100\,per\,cent. We assume this extends to fainter radio source identifications in the 6dFGS-NVSS sample.
Multiple-component radio source identifications are rare (they comprise $<1$\,per\,cent of
all radio sources in this sample) and we believe these cannot affect the completeness and reliability
in a significant way. The missing population of GRGs described in Section~\ref{GRGs} also contributes
$<1$\,per\,cent to the incompleteness of 6dFGS-NVSS identifications.

In summary, we find that the completeness of the 6dFGS-NVSS database is better than 96\,per\,cent (a 3\,per\,cent
contribution from missing single-component matches and a $\sim1$\,per\,cent contribution from missing multiple-component matches). The
reliability of the database is better than 99\,per\,cent. Recently \citet{best05} described a robust identification
method utilising both the FIRST and NVSS surveys to find radio sources in the SDSS and their
matching procedure had an estimated completeness of 94.4\,per\,cent and reliability of 98.9\,per\,cent. The FIRST
survey only covers about 1\,per\,cent of the region observed in the 6dFGS\,DR2 and we were therefore unable
to use their method. We achieve similar (or slightly better) completeness and reliability to them by visually
classifying all radio-optical identifications; we believe this is the most robust method of identifying
radio sources in optical sky surveys when only the NVSS is available.

\subsection{Classification of 6dFGS optical spectra}
\label{specclass}

\begin{figure*}
\centering
\includegraphics[width=8.7cm]{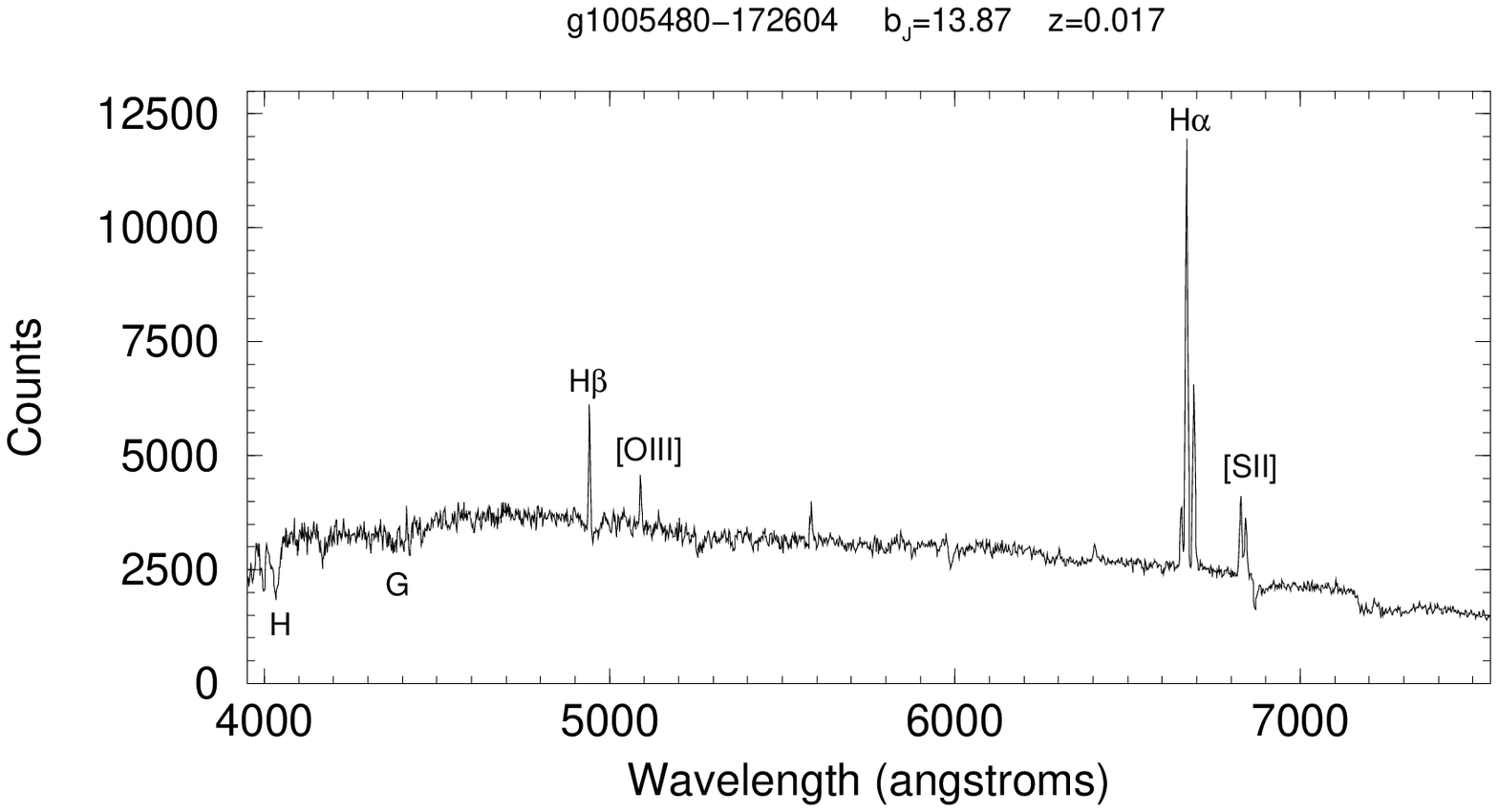}%
\includegraphics[width=8.7cm]{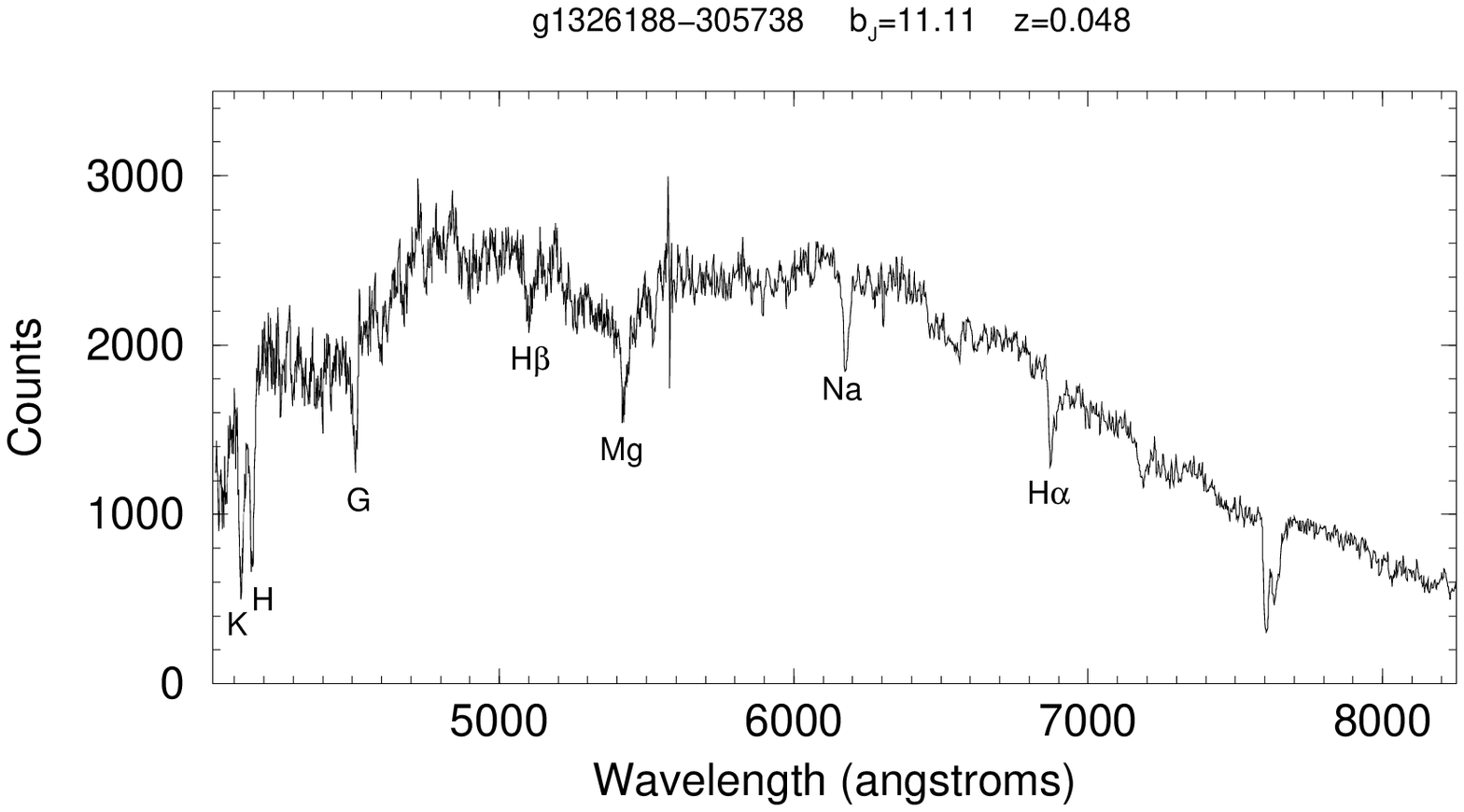}
\includegraphics[width=8.7cm]{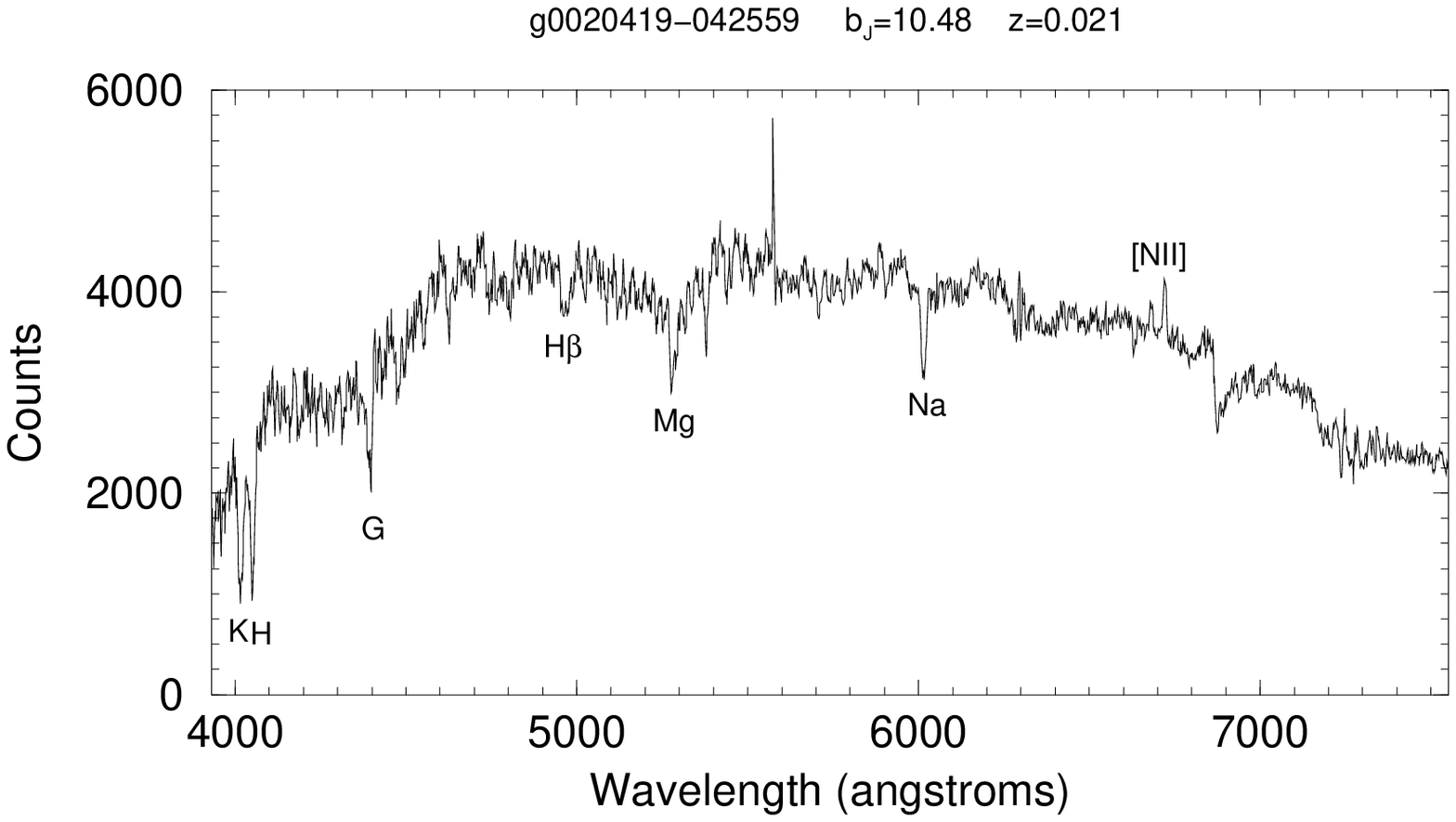}%
\includegraphics[width=8.7cm]{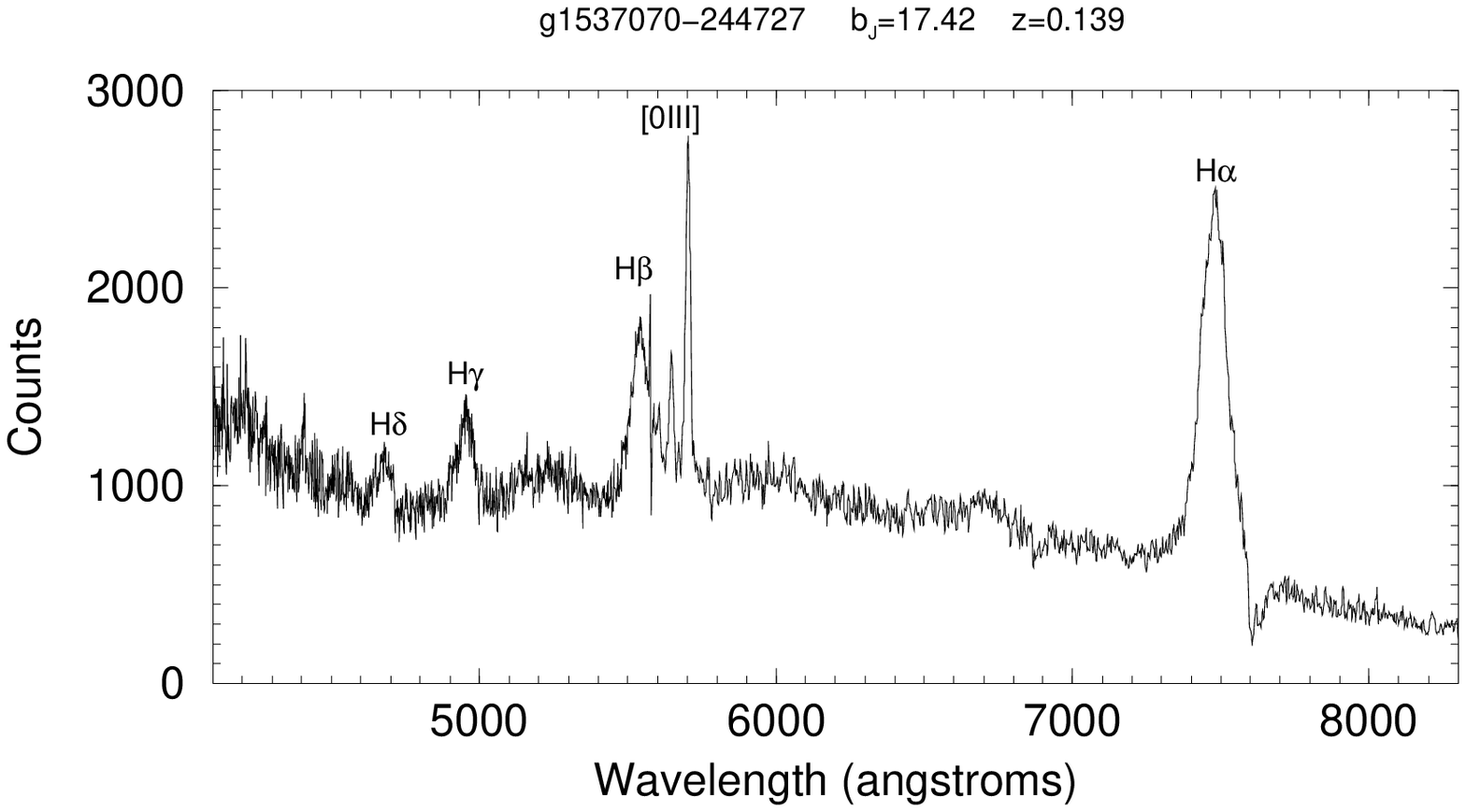}
\caption{Example 6dFGS spectra 
  of 4 spectral classes. In each spectrum common redshifted
  absorption/emission
  lines are labelled at the measured redshift.
  Top left: An `SF' galaxy, with characteristic
  strong and narrow Balmer lines. Top right: An `Aa' galaxy, containing
  broad Mg absorption as well as a distinct break close to the H \& K lines. 
  Bottom Left: An `Aae' galaxy. This looks similar to an `Aa'
  galaxy but has a weak [N{\sc ii}] line. Bottom right: 
  An `Ae' galaxy. This has
  broad H$\alpha$ emission and strong [O{\sc III}] lines.}
\label{specclassexamp}
\end{figure*}

\begin{table}
\centering
\caption{Spectral classes visually
assigned to 6dFGS-NVSS objects.}
\label{claspec}
\begin{tabular}{rlc}
\hline
Class & Type of spectrum & Number\\
\hline
Aa & Pure absorption line spectrum. & 2124 \\
Aae & Narrow LINER-like emission lines. & 370 \\
Ae & Emission line AGN. & 370\\
SF & Star-forming galaxy. & 4625\\ 
? & Unclassifiable or unknown. & 321 \\
star & Galactic star. & 14\\
\hline
Total & & 7824\\
\hline
\end{tabular}
\end{table}

The 6dF spectrum of each accepted radio source
identification was inspected visually to 
determine the dominant physical process responsible
for the radio emission of each galaxy. 
We followed the scheme described by \citet{2dfnvss} and classified each spectrum as
either `AGN' for spectra indicative of galaxies harbouring a radio-loud active
galactic nucleus or `SF' for spectra indicative of galaxies with ongoing
star formation. 
AGNs can have a pure absorption line spectrum like
that of a giant elliptical galaxy (classed as Aa); a spectrum with
absorption lines and weak narrow LINER-like emission lines (classed as
Aae); a conventional Type II AGN
spectrum which has nebular emission lines such as [O{\sc ii}]
, [O{\sc iii}] or [N{\sc ii}] 
which are stronger than any hydrogen Balmer emission
lines  (H$\alpha$ or H$\beta$) (classed as Ae); or a conventional Type I
AGN spectrum with strong
and broad hydrogen Balmer emission lines (also classed as Ae). 
Star-forming galaxies have
spectra typical of H{\sc ii} regions with strong narrow emission lines of
H$\alpha$ and H$\beta$ dominating the spectrum (classed as SF). Figure~\ref{specclassexamp}
shows an example of each type of spectrum. Spectra of galactic stars were
classified `star' and spectra for which a classification could not be made
(mostly due to a low S/N spectrum)
were classed `?'. A `?' was also appended to the end of any classification
which was not certain. Our visual classification scheme is summarised 
in Table~\ref{claspec}.

To check the reliability of the visual classification scheme a subsample
of $\sim1800$ spectra was classified by both of us and the classifications
compared. The two human classifiers agreed well, with
disagreement in about 10\,per\,cent of cases. The primary cause of disagreement is
between the classification of emission-line AGN (Aae and Ae classes) and
the SF class, and occurs for spectra which have line
ratios which are borderline between the AGN and SF classes.
If only
the AGN and SF classes are considered then the two human classifiers
disagree only 5\,per\,cent of the time. 
In their analysis of 2dFGS spectra using the same classification scheme
described here,
\citet{2dfnvss} found that this scheme agreed well with
methods such as principal component analysis 
\citep[PCA;][]{folkes99,madgwick02} and 
classification based upon 
diagnostic emission line ratios~\citep{jackson00} both of which have 
a similar reliability to that of the visual classification
presented here.

\section{Sample Properties}
\label{primsamplesec}

We matched a total of 7\,824 6dFGS\,DR2 primary targets with radio
sources from the NVSS catalogue, corresponding to a radio detection
rate of 16.5\,per\,cent. Of the 7\,824 detections, 7\,672 were identified
with a single NVSS component and 152 were identified with more than one. 
Table~\ref{claspec}
summarises the spectral properties of the accepted 6dFGS-NVSS matches.
Out of the 7\,824 spectra we examined, 321 were unclassifiable (classed as '?')
and 14 of them were Galactic stars 
(assigned redshift quality $Q=6$ in the 6dFGS).
Of the unclassifiable spectra, 302 had redshift quality $Q=1$ or $2$, these
spectra were too noisy for reliable spectral classification.
The remaining 19 unclassifiable spectra had reliable redshift measurements, but
did not present enough information for reliable visual classification.

The ratio of star-forming galaxies (SF) to radio-loud AGN in the sample
is approximately 60\,per\,cent SF to 40\,per\,cent 
radio-loud AGN. The fractions of SF to 
radio-loud AGN in the similarly classified 2dFGRS-NVSS sample \citep{2dfnvss}
are 40\,per\,cent SF to 60\,per\,cent 
radio-loud AGN. The difference in these relative
fractions can be
explained by the fainter magnitude limit ($b_J=19.4$) of the 2dFGRS,
resulting in a larger proportion of detections of more distant radio-loud 
AGN. It is interesting
that in both the 6dFGS-NVSS and 2dFGRS-NVSS samples, over 70\,per\,cent of all radio-loud AGN
have absorption-line spectra and would be missed from AGN 
samples selected on the basis of optical emission lines.

\subsection{The data table}

\begin{table*}
\begin{center}
\caption{The first 30 entries of the 6dFGS-NVSS data table.}
\label{datatable}
\begin{tabular}{ccccccccccc}
\hline
(1) & (2) & (3) & (4) & (5) & (6) & (7) & (8) & (9) & (10) & (11) \\
 6dFGS target & RA & Dec. & $K$ & $z$ & $Q$ & Offset & $S_{\rm 1.4\,GHz}$ & $S_{\rm 843\,MHz}$ & $S_{\rm 60\,\mu m}$ & Spectrum \\ 
 name & \multicolumn{2}{c}{(J2000)} & & & & & NVSS & SUMSS & IRAS\,FSC & Class \\
      & $(h : m : s)$ & $(^{\circ}\,:\,\arcmin\,:\,\arcsec)$ & (mag.) & & & (arcsec) & (mJy) & (mJy) & (Jy) & \\ 
\hline
g0000124$-$363113 & 00:00:12.39 & $-$36:31:13.1 & 12.724 & 0.1169 & 4 &  9.5 &   12.1 &  &  &   Aa \\
g0000141$-$251113 & 00:00:14.12 & $-$25:11:12.9 & 11.886 & 0.0852 & 4 &  3.8 &   28.4 &  &  &   Aa \\
g0000356$-$014547 & 00:00:35.64 & $-$01:45:47.4 & 11.267 & 0.0246 & 4 & 15.4 &    2.8 &  &  & Aae? \\
g0000523$-$355037 & 00:00:52.33 & $-$35:50:37.2 & 11.330 & 0.0521 & 4 &  0.0 &   48.4 &  &  &   Aa \\
g0001197$-$140423 & 00:01:19.74 & $-$14:04:23.2 & 12.289 & 0.0867 & 3 & 11.0 &   40.5 &  &  & Aae? \\
g0001394$-$025852 & 00:01:39.36 & $-$02:58:52.1 & 12.001 & 0.1013 & 4 &  2.3 &    5.5 &  &  &   Aa \\
g0001453$-$042049 & 00:01:45.29 & $-$04:20:49.0 & 11.845 & 0.0481 & 4 &  5.0 &    2.8 &  &  &   Aa \\
g0001496$-$094138 & 00:01:49.58 & $-$09:41:37.8 & 12.304 & 0.1038 & 4 & 13.8 &    3.4 &  &  &   Aa \\
g0001558$-$273738 & 00:01:55.82 & $-$27:37:38.0 & 10.164 & 0.0283 & 4 &  4.3 &   29.8 &  &   0.488 &   SF \\
g0001567$-$035755 & 00:01:56.70 & $-$03:57:54.8 & 12.492 & 0.0226 & 4 &  2.3 &    6.2 &  &   0.466 &   SF \\
g0001572$-$383857 & 00:01:57.20 & $-$38:38:56.4 & 12.644 & 0.0547 & 4 &  2.5 &   11.8 &    22.9 &  &   SF \\
g0002039$-$332802 & 00:02:03.88 & $-$33:28:02.2 & 11.143 & 0.0289 & 4 &  4.2 &    6.2 &  &   0.941 &   SF \\
g0002348$-$034239 & 00:02:34.81 & $-$03:42:38.6 & 10.623 & 0.0215 & 4 &  0.3 &   13.1 &  &   1.121 &   SF \\
g0002487$-$033622 & 00:02:48.65 & $-$03:36:21.7 & 11.455 & 0.0208 & 4 &  6.9 &    3.9 &  &  &   SF \\
g0002547$-$354319 & 00:02:54.69 & $-$35:43:19.4 & 11.677 & 0.0489 & 4 & 24.0 &    4.2 &  &  &   Aa \\
g0002545$-$341408 & 00:02:54.47 & $-$34:14:08.4 & 10.699 & 0.0226 & 4 &  2.6 &    7.3 &  &   0.760 &   SF \\
g0002558$-$265451 & 00:02:55.81 & $-$26:54:51.2 & 11.994 & 0.0665 & 4 &  1.1 &   94.4 &  &  &   Aa \\
g0003051$-$073700 & 00:03:05.06 & $-$07:37:00.3 & 12.291 & 0.0299 & 4 &  6.3 &    3.7 &  &  &  Aae \\
g0003057$-$015450 & 00:03:05.66 & $-$01:54:49.7 & 10.156 & 0.0244 & 4 &  1.1 &    7.7 &  &   0.805 &   SF \\
g0003056$-$295159 & 00:03:05.62 & $-$29:51:59.3 & 11.741 & 0.0609 & 4 &  6.7 &   27.8 &  &  &   Aa \\
g0003130$-$355614 & 00:03:12.97 & $-$35:56:13.4 & 10.508 & 0.0499 & 4 &  1.4 &  589.5 & 1123.4 &  &   Aa \\
g0003321$-$104441 & 00:03:32.13 & $-$10:44:40.6 & 10.348 & 0.0299 & 4 &  8.0 &    2.5 &  &  & Aae? \\
g0003449$-$204757 & 00:03:44.90 & $-$20:47:56.5 & 12.314 & 0.0970 & 4 & 18.8 &    3.2 &  &   0.274 &   SF \\
g0004029$-$330202 & 00:04:02.86 & $-$33:02:02.0 & 12.397 & 0.0380 & 4 & 12.9 &    2.4 &  &   0.447 &  SF? \\
g0004472$-$013413 & 00:04:47.22 & $-$01:34:12.8 & 11.831 & 0.0239 & 4 &  5.1 &    9.9 &  &   0.843 &   SF \\
g0004517$-$060058 & 00:04:51.74 & $-$06:00:57.6 & 12.722 & 0.1080 & 4 &  1.5 &    9.3 &  &  &   Aa \\
g0004576$-$014108 & 00:04:57.57 & $-$01:41:07.9 & 12.569 & 0.0239 & 4 &  9.3 &    3.6 &  &  &   SF \\
g0005026$-$160715 & 00:05:02.58 & $-$16:07:15.0 & 11.997 & 0.0339 & 4 & 13.8 &    2.2 &  &   0.268 &   SF \\
g0005028$-$274253 & 00:05:02.78 & $-$27:42:52.5 & 11.619 & 0.0333 & 4 &  2.9 &   10.2 &  &   1.178 &   SF \\
g0005054$-$070536 & 00:05:05.37 & $-$07:05:36.3 & 10.065 & 0.0128 & 4 &  2.6 &   12.9 &  &   1.038 &   SF \\
\hline
\end{tabular}
\end{center}
\begin{flushleft}
\footnotesize
Column descriptions:\\
(1) The target name of the object from the 6dFGS database.\\
(2) \& (3) J2000 Right Ascension and declination of the object from the 2MASS\,XSC.\\
(4) Total K-band magnitude calculated from 2MASS isophotal K magnitude according to equation 1 of \citet{jones04}.\\
(5) \& (6) 6dFGS measured redshift and quality flag as described in \citet{jones04}. Only redshifts with ${\rm Q}\geq3$ were
deemed reliable.\\
(7) The offset in arcsec from the NVSS radio position to the objects position in the 6dFGS database.\\
(8) The 1.4\,GHz flux density in mJy from the NVSS catalogue.\\
(9) The 843\,MHz flux density in mJy for 6dFGS objects with a SUMSS catalogue match within 10\,arcsec.\\
(10) The IRAS\,FSC $60\,\mu{\rm m}$ flux density in Jy for 6dFGS objects which also appear in additional target sample 126 
\citep[see][]{jones04} of the 6dFGS.\\
(11) The classification of the spectrum as defined in Table~\ref{claspec}.\\
\end{flushleft}
\end{table*}

Table~\ref{datatable} shows details of the sample of 6dFGS-NVSS 
radio sources. The 6dFGS target name is given for each source so that
the observed 6dFGS spectra can be obtained or each object can be matched
against other 6dFGS additional target samples \citep{jones04}. The table
also shows 2MASS\,XSC J2000 positions, derived $K_{\rm tot}$, redshifts
and redshift quality $Q$ for each source. We list offsets 
from the given position and integrated 1.4\,GHz flux densities 
for each NVSS radio source. For single component NVSS radio sources with 
an 843\,MHz SUMSS catalogue (version 1.6) detection within 10\,arcsec we also list
the 843\,MHz flux density of the source. The list of SUMSS matches is highly
incomplete at present as the 6dFGS DR2 covers little of the sky
south of $\delta=-40^\circ$ and version 1.6 of the SUMSS catalogue
is not complete north of this declination. We plan to thoroughly investigate the
properties of SUMSS radio sources in the 6dFGS when the full data releases 
of both surveys are made available and this will be the subject of a future paper.
For objects with a match in the 6dFGS additional target sample selected from
the Infrared Astronomical Satellite (IRAS) Faint Source Catalogue (FSC) 
\citep{irasfsc} \citep[programme id 126;][]{jones04} 
we list the $60\,\mu$m flux density
of the source. The final column of the table shows the visually assigned 
spectral classification described in Section~\ref{specclass}.

\subsection{Radio source counts}

\begin{figure}
\centering
\includegraphics[width=\linewidth]{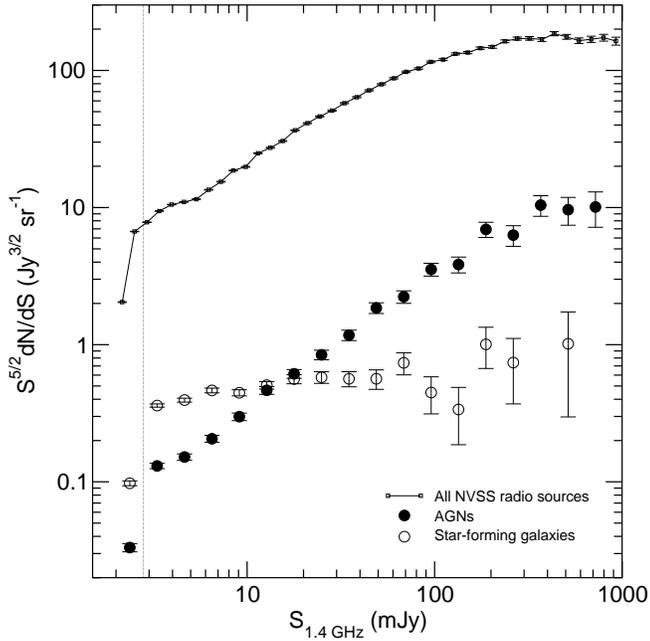}
\caption{ Differential source counts of 1.4\,GHz NVSS sources multiplied by
  $S^{\frac{5}{2}}$ such that counts in a Euclidean universe would lie on a
  straight line. The source counts of all 580\,419 NVSS sources in the region
  surveyed are shown by open squares linked by a solid line. AGN and SF
  class galaxies in the 6dFGS-NVSS sample are shown as filled and open
  circles respectively. The error bars shown are the $\sqrt{n}$ 
  counting uncertainties in each bin. A vertical line is plotted at 
  $S_{1.4\,{\rm GHz}}=2.8$\,mJy which is the completeness limit of the 6dFGS-NVSS sample
	and is also the upper bound of the faintest bins calculated from the AGN and SF samples.}
\label{primsc}
\end{figure}

The differential
source counts of 6dFGS-NVSS galaxies are plotted in
Figure~\ref{primsc}. These are plotted in a form in which each bin is
weighted by $S^{5/2}$ such that counts in a static Euclidean universe would
lie on a horizontal line. 
Counts determined for the radio-loud AGN are plotted as
filled circles and those for star-forming galaxies are plotted as open
circles. The source counts of all 580\,419 NVSS catalogue sources in the 
same region of sky as the 6dFGS-NVSS sample are shown as squares linked
by a solid line. The counts follow an approximate power law
between 1\,Jy and 2.8\,mJy, below which they fall sharply; this is because
the NVSS catalogue becomes increasingly 
incomplete below 2.8\,mJy \citep{nvss}. The vertical line at 2.8\,mJy 
shown in the figure divides the two smallest bins of the AGN and SF samples.
The same falloff can be seen to the left of the vertical line
in the counts of the AGN and SF subsamples indicating
that this 6dFGS-NVSS sample is complete to a flux density limit of 2.8\,mJy.

Radio source catalogues such as the NVSS are known to predominantly contain
galaxies powered by radio-loud AGN with a median
redshift of $\tilde{z}\sim0.8$ \citep{nvss}. In the much more nearby 
($\tilde{z}\sim0.05$) 6dFGS-NVSS sample the source counts of radio-loud AGN 
follow roughly the same
power-law slope as that of all radio sources but only account
for roughly 1\,per\,cent of their total number.
The source counts of star-forming galaxies
stay flat over the range 2.8\,mJy to 1\,Jy and
therefore account for only 0.1\,per\,cent of all sources at 1\,Jy rising to 
1-2\,per\,cent of
all sources at the survey limit. The counts of star-forming galaxies and 
radio-loud AGN 
cross over at about 10\,mJy, below which star-forming galaxies dominate the
population of radio sources. At 1\,Jy there are
over ten times more radio-loud AGN than star-formers, but the
balance changes at lower flux density 
until at 2.8\,mJy there are nearly ten times
more star-formers than radio-loud AGN in this volume-limited sample.

\subsection{Redshift distribution}

\begin{figure}
\centering
\includegraphics[width=\linewidth]{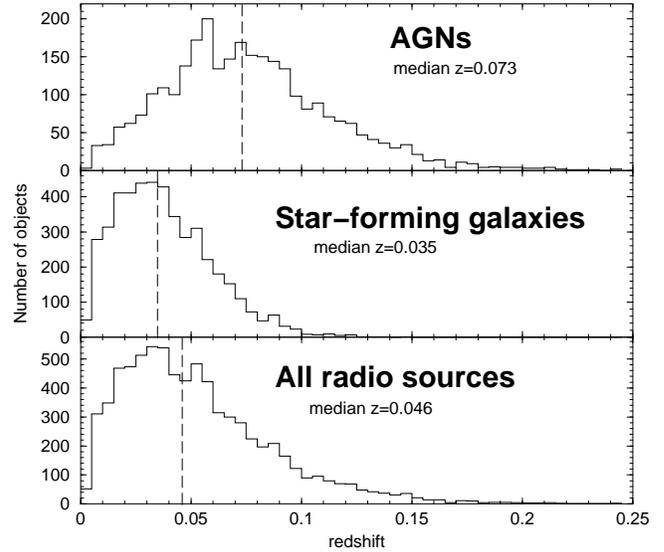}
\caption{The redshift distribution of the 6dFGS-NVSS sample in bins
  of width 0.005 in redshift. 
  \textit{Bottom panel}: Redshift distribution of all 7\,824 objects for which 
  a redshift has been measured. \textit{Middle panel}: Redshift
  distribution of the 4\,625 Star-forming galaxies. \textit{Top panel}: 
  Redshift distribution of the 2\,864
  radio-loud AGN. In each panel a vertical dashed line is drawn at the median
  redshift.}
\label{primzdist}
\end{figure}

\begin{figure}
\centering
\includegraphics[width=\linewidth]{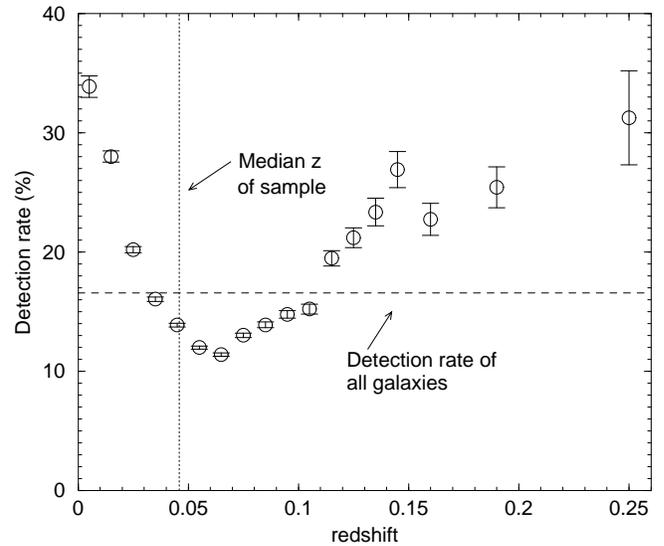}
\caption{ The NVSS detection rate of 6dFGS\,DR2 primary targets. 
Here the detection rate is defined as the
  percentage of radio identifications of
  6dFGS DR2 objects for that bin. The error bars shown
  correspond to 
  the $\sqrt{n}$ counting error in each bin. A vertical dotted line is
  plotted at the median redshift of the 6dFGS-NVSS sample and a horizontal
  dashed line is plotted at the overall radio detection rate of 16.5\,per\,cent.}
\label{detratez}
\end{figure}

Fig.~\ref{primzdist} shows the redshift distribution of 6dFGS-NVSS objects. 
The redshift distribution of all objects is shown in
the bottom panel and separate distributions for star-forming galaxies and 
radio-loud AGN
are shown in the middle and top panels respectively. The median redshift
of all radio sources in the 6dFGS-NVSS sample is $\tilde{z}=0.046$, very
close to the median redshift ($\tilde{z}=0.054$) 
of the underlying $K$-selected sample
\citep{jones04}. Star-forming galaxies dominate the lower
redshift radio source population and are found at a median redshift of
$\tilde{z}=0.035$ while radio-loud AGN dominate the higher redshift radio source
population at a median redshift of $\tilde{z}=0.073$. 

The optical magnitude limits, radio flux density limits and median
redshifts of other radio-optical samples, selected in a similar way
to the 6dFGS-NVSS 
sample, are presented in Table~\ref{samplecompare}.
In their analysis of $S_{1.4\,{\rm GHz}}\geq2.8$\,mJy 2dFGRS-NVSS radio sources  
\citet{sadler99} found
a median redshift of $\tilde{z}=0.1$ for all 2dFGRS-NVSS galaxies, 
$\tilde{z}=0.05$ for
star-forming galaxies and $\tilde{z}=0.14$ for radio-loud AGN. 
The 2dFGRS goes more than two magnitudes deeper in $b_J$ than the 6dFGS yet
the median redshift
of 2dFGRS-NVSS star-forming galaxies is only slightly higher than that found
here for 6dFGS-NVSS galaxies, since it is primarily the radio
flux density limit of optical-radio samples which limits the maximum distance
to which star-forming galaxies can be found. In a survey of 2dFGRS-FIRST radio
sources with a fainter flux density limit of 
$S_{1.4\,{\rm GHz}}\geq1$\,mJy, \citet{maglio02} found a
median redshift for
star-forming galaxies of $\tilde{z}=0.1$, 
more than twice that of the NVSS selected
samples. Conversely, for radio-loud AGN the median redshifts of the
2dFGRS-NVSS and 2dFGRS-FIRST samples are roughly similar ($\tilde{z}=0.15$)
whereas the median redshift of the 6dFGS-NVSS sample is significantly
lower. This is a consequence of the different optical magnitude limits
of the 6dFGS and the 2dFGRS; radio-loud AGN fall out of both samples at
the optical/near-infrared limit of the spectroscopy.

\subsection{Detection rates}

Fig.~\ref{detratez} shows the variation in the detection rate of 
6dFGS-NVSS radio sources as a function of
redshift. The detection rates were calculated
by dividing the number of 6dFGS-NVSS galaxies in redshift 
bins of width 0.01 by 
the number of the 6dFGS DR2 
primary targets with $Q=3$ or $4$ in each redshift bin. 
The average NVSS detection rate is shown as a dashed
horizontal line in and the median redshift of the 6dFGS-NVSS
sample is shown as a dotted vertical line. The NVSS detection rate drops
sharply from 34\,per\,cent at $z=0.005$ to just over 11\,per\,cent 
at $z=0.05$ (around the median
redshift of the survey). It then rises steadily to over 
20\,per\,cent at $z=0.15$
beyond which results are dominated by counting errors. This fall and
subsequent rise in the detection rate with redshift can be explained in
terms of the different redshift distributions 
of the star-forming and radio-loud AGN populations. 

Below $z=0.05$, star-forming galaxies dominate the radio source population
and the detection rate is falling because of the 1.4\,GHz flux density limit
of the NVSS. Star-forming galaxies have weaker radio
powers and fewer are found with increasing redshift because the limiting radio
power of the 6dFGS-NVSS sample is increasing with redshift. 
Above $z=0.05$, the detection rate is increasing and
radio-loud AGN
dominate the radio source population. The population of radio-loud AGN has
a higher average radio power so they tend not to fall out of the
sample because of the flux density limit of the NVSS. 
The $K\leq12.75$\,mag. cutoff of the
6dFGS means that as redshift increases the galaxies left
in the survey are on average more luminous in the near-infrared.
Radio-loud AGN preferentially reside within these most luminous galaxies
(eg. Fig.~\ref{primlumdist}), 
resulting in an increase in the detection rate of radio-loud AGN.

\subsection{Luminosity distribution}
\label{primlumsec}

\begin{figure}
\centering
\includegraphics[width=\linewidth]{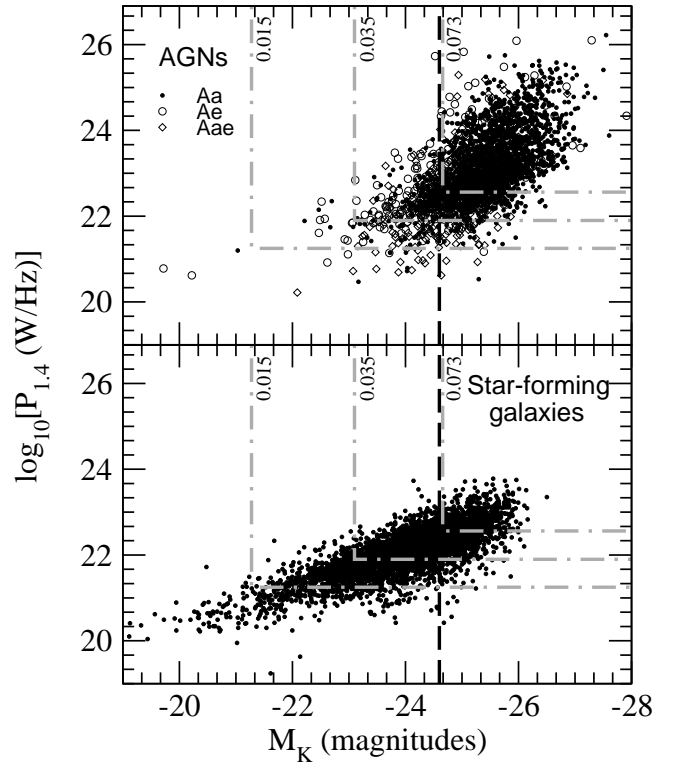}
\caption{\textit{Bottom}: 1.4\,GHz radio power vs. $M_K$ for the 4\,625
  star-forming galaxies in the 6dFGS-NVSS sample.
  \textit{Top}: 1.4\,GHz radio power vs. $M_K$ for the
  2864 radio-loud AGN in the 6dFGS-NVSS sample. Aa class AGN are plotted as dots,
  Aae class AGN are plotted as circles and Ae class AGN are plotted as 
  diamonds. 
  A dashed vertical line is drawn at the value of $M_K^\star=-24.6$\,mag.
  from a Schechter function fit to the local $K$-band luminosity function of galaxies
  in the 6dFGS by \citet{jones06}. The grey dot-dashed lines in each panel show the
  limiting values of $P_{1.4}$ and $M_K$ derived from the radio ($S_{1.4\,{\rm GHz}}=2.8$\,mJy) 
  and near-infrared ($K=12.75$\,mag.) completeness limits of the 6dFGS-NVSS sample
  at the labelled redshifts (ie. $z=0.015$, $z=0.035$, the median redshift
  of star-forming galaxies \& $z=0.073$, the median redshift of radio-loud AGN). These
  lines show that the apparent correlation between $P_{1.4}$ and $M_K$ is caused by the 
  flux density and apparent magnitude limits and the range in redshift of the sample.}
\label{primlumdist}
\end{figure}

Fig.~\ref{primlumdist} shows
the 1.4\,GHz monochromatic radio power ($P_{1.4}$, in units of W\,Hz$^{-1}$)
vs. the absolute K-band magnitude ($M_{K}$) for the radio-loud AGN (\textit{top})
and star-forming galaxies (\textit{bottom}). 
Radio power measurements have been $k$-corrected 
using the usual form, $k_{\rm{radio}}(z)=(1+z)^{-(1+\alpha)}$
where $\alpha$ is the spectral index ($S_{\nu} \propto \nu^{\alpha}$). 
We determined median spectral indices
separately for the AGN and SF classes by comparing the
flux densities of the 436 galaxies that were also found in version
1.6 of the 843\,MHz SUMSS catalogue \citep{sumss}. For the 228 
radio-loud AGN the median spectral index was
$\tilde{\alpha}^{1400}_{843}=-0.54\pm0.07$ and for the 208 SF galaxies the median
  spectral index was $\tilde{\alpha}^{1400}_{843}=-0.90\pm0.07$. These values
  of $\alpha$ have been applied to the $k$-corrections 
  for all calculations of radio power
  throughout this paper. Both these
  values are close to the value of $\alpha=-0.7$ often assumed for radio
  source samples \citep[eg.][]{2dfnvss,condon02}. For most of the
  galaxies in this sample $k_{\rm{radio}}(z)$ is negligibly small.

In the $K$-band,
$k$-corrections ($k_K(z)$) are similar for galaxies of all Hubble types
because different amounts of star formation in different galaxy types only
significantly affect galaxy spectra at wavelengths shorter than $1\,\mu\rm{m}$
\citep{cole01,glazebrook95}. $K$-band $k$-corrections
have been calculated following \citet{glazebrook95} 
who derived $k_K(z)$ from the evolutionary synthesis models
of \citet{kc93} assuming an instantaneous burst of star formation at age
5\,Gyr. 
At 
the small redshifts probed by this dataset $(z<0.2)$, $k_K(z)$
never changes $M_K$ by more than 0.4 magnitudes.

The star-forming galaxies in Fig.~\ref{primlumdist} span a range
of $M_K$ between $-20$ and $-26$, they have median $\tilde{M_K}=-24.13$,
lower than $M_K^\star$. Star-forming galaxies, 
with median $\log[\tilde{P}_{1.4}$\,(W\,Hz$^{-1})]=22.13$,
have weaker radio powers ($P_{1.4}<10^{23}$\,W\,Hz$^{-1}$)
than radio-loud AGN. 
Radio-loud AGN are almost all
found in objects brighter than $M_K^\star$, indicative of their preferential
location in the brightest galaxies. The median $\tilde{M_K}$ of
radio-loud AGN is $-25.40$ and they have 
median radio power $\log[\tilde{P}_{1.4}$\,(W\,Hz$^{-1})]=23.04$, almost an
order of magnitude brighter than the median radio power of star-forming
galaxies. Though star-forming galaxies have lower radio powers, 
radio-loud AGN span a wide range in radio
power from $10^{21}$ to $10^{26}$\,W\,Hz$^{-1}$, indicating that there is
no clear observational regime in which star-forming galaxies can be
separated from radio-loud AGN purely on the basis of radio power. 

\section{The FIR-radio correlation}
\label{primiras}

It has been well established that a correlation exists between
far-infrared (FIR) and radio continuum emission from normal galaxies
(see \citet{helou85,cb98,condon91,condon92}). 
This correlation has been attributed to ongoing
star formation within the host galaxy, where radio continuum emission is
produced by a combination of 
synchrotron emission from electrons accelerated in the 
supernova remnants of short-lived massive stars ($M>8M_\odot$ with
lifetimes $<3\times10^7$\,yr) and free-free emission from H{\sc ii} regions
ionised by these same massive stars \citep{condon92}. 
FIR emission is caused by thermal
reradiation of dust in H{\sc ii} regions heated by this same
population of massive stars. Though it is understood that recent 
star formation is the process which drives both the radio continuum and FIR
emission of galaxies which lie on the correlation, the actual mechanism
that relates non-thermally dominated radio power and thermally dominated
FIR luminosity is poorly understood. 

The FIR-radio correlation has served as a diagnostic tool for large samples
of radio sources, as it can be used to distinguish between galaxies with
ongoing star formation and those harbouring a radio-loud AGN. 
Galaxies falling on the correlation derive their radio emission from star
formation, whereas galaxies with a radio
``excess'' above that expected from the correlation derive their radio
emission from the presence of a radio-loud AGN \citep[eg.][]{condon02,cb98}. 
The correlation also 
constrains models relating star-formation rate to radio emission
\citep{condon92}. 
In this section results from
analysis of the population of radio sources in the 6dFGS-NVSS
sample with FIR detections in the IRAS\,FSC
are presented, both as a consistency check for
the spectral classification of the sample and to examine the correlation
for a large sample of normal galaxies.

\subsection{Finding IRAS Detections}

In addition to the primary $K$-selected sample, the 
6dFGS aims to measure redshifts for several additional target samples 
selected at various wavelengths.
One of these additional target lists consists of
$\sim 11\,000$ IRAS\,FSC galaxies selected at $60\,\mu{\rm m}$ from
the IRAS\,FSC
\citep[For a complete description of the 6dFGS additional target
samples see][]{jones04}. 
Much of this IRAS\,FSC additional target sample (which we call the 6dFGS-FSC sample)
overlaps with our 6dFGS-NVSS sample. It is
therefore straightforward to find IRAS-FSC detections of
6dFGS-NVSS primary target objects by crossmatching 
their 6dF target ID names.
4\,403 6dFGS-FSC targets were observed
in the DR2. Table~\ref{irasdets} shows the spectral
classification of the 2\,942 galaxies which are common to the 6dFGS-NVSS
sample and the 6dFGS-FSC sample.
Predominantly emission line (ie. Ae, Aae, SF) galaxies are
detected in the far-infrared as has been observed in other
spectroscopic studies of far-infrared selected galaxies
\citep[eg.][]{lawrence86,grijp92,2dfnvss}. The four Aa galaxies
detected in the 6dFGS-NVSS-FSC subsample have low S/N and 
may have been misclassified.

\begin{table}
\begin{center}
\caption{Spectral classification of 6dFGS-NVSS objects which are also found
  in the 6dFGS-FSC sample.}
\label{irasdets}
\begin{tabular}{lrr}
\hline
Spectral class & 
${N^{\rm 6dFGS-FSC}_{\rm 6dFGS-NVSS}}^1$ & per\,cent of total$^2$\\
\hline
SF & 2690 & 58.1 \\
Ae & 138 & 37.3   \\
Aae & 66 & 17.8   \\
Aa  &  4 &  0.2    \\
star & 2 & 14.3    \\
? & 42 &  13.1     \\
Total & 2942 & 37.6 \\
\hline
\end{tabular}
\end{center}
\footnotesize
NOTES:\\
$^1$ The number of 6dFGS DR2 objects which are common to both the IRAS\,FSC and the
NVSS catalogue.\\
$^2$ The percentage of 6dFGS-NVSS objects that are also 6dFGS-FSC objects.\\
\end{table}

\citet{condon02} noted that the FIR-radio correlation ensures that most 
far-infrared
sources powered by ongoing star formation are also radio sources and
vice versa. The FIR/radio ratio 
$u=\log(S_{60\,\mu{\rm m}}/S_{1.4\,{\rm GHz}})$ for nearby spiral
galaxies in the UGC has a mean value of $\left<u\right>=2.05\pm0.02$ with
rms width $\sigma_u=0.2$ \citep{cb98}. This value of $\left<u\right>$
happens to be the ratio between the 280\,mJy flux density
limit of the IRAS-FSC at $60\,\mu$m and the limit of
the NVSS at 1.4\,GHz $(u \approx 2.05)$. 
58.1\,per\,cent of spectroscopically classified SF galaxies
have 6dFGS-FSC detections leaving 929 SF galaxies with no FIR
detection. These missing detections are the result of incompleteness
at the flux density limits of the 6dFGS-FSC and 6dFGS-NVSS samples and 
incomplete sky coverage of 
the region $-40^\circ<\delta<0^\circ$ in the IRAS\,FSC 
\citep{irasfsc}.

When available, $100\,\mu$m flux densities from the IRAS\,FSC
were also obtained for each 6dFGS-FSC source. For the $\sim\,20$\,per\,cent of objects which did
not have a $100\,\mu$m detection the $100\,\mu$m flux density was assumed
to be twice the $60\,\mu$m flux density based on the calculated average
$\left<\log(S_{100\,\mu{\rm m}}/S_{60\,\mu{\rm m}})\right>=0.3$ with a
scatter of 0.2 from the bright galaxy 
sample of \citet{soifer89}. 60 and $100\,\mu{\rm
  m}$ flux densities are then converted into the quantity $FIR$ (in units of
W\,m$^{-2}$) defined by
\begin{equation}
\label{fireqn}
{\rm FIR}=1.26\times10^{-14}\left(2.58S_{60\,\mu{\rm m}} + S_{100\,\mu{\rm
      m}}\right)
\end{equation}
where $S_{60\,\mu{\rm m}}$ and $S_{100\,\mu{\rm m}}$ are in Janskys
\citep{helou85}. $FIR$ is a measure of the total far-infrared flux between
42.5 and 122.5\,$\mu$m. 

\subsection{The FIR-radio ratio $q$}

The radio-FIR correlation is often parametrised by the FIR-radio flux ratio
parameter $q$ \citep{helou85}. This is defined by
\begin{equation}
\label{q}
q=\log\left(\frac{{\rm FIR}/(3.75\times10^{12})}{S_{1.4\,{\rm GHz}}}\right)
\end{equation}
where ${\rm FIR}$ (defined by equation~\ref{fireqn}) is divided by the factor
$3.75\times10^{12}$\,Hz (the frequency at $80\,\mu$m) to convert ${\rm FIR}$ to
W\,m$^{-2}$\,Hz$^{-1}$.
The mean value of
$\left<q\right>=2.28$ with rms scatter $\sigma_q=0.22$ 
for all 2\,411 objects with measured flux densities at both 60 and
100\,$\mu$m. $\sigma_q$ is somewhat larger than the value found by
\citet{condon02} for a more nearby sample of UGC-NVSS galaxies. 
This discrepancy
reflects the large uncertainties associated with fainter
IRAS\,FSC flux 
densities as well as the larger radio-loud AGN population contained in the
present sample. For the subset of 2\,242
star-forming galaxies with measured 60 and $100\,\mu$m flux densities
$\left<q_{\rm SF}\right>=2.3$ with rms scatter
$\sigma_{\rm SF}=0.18$, in close agreement with \citet{condon02} for the
UGC-NVSS sample. This value also agrees well with 
results from samples of stronger IRAS sources in spiral
galaxies~\citep{condon91}. For the subset of 169 radio-loud AGN with measured 60 and
$100\,\mu$m flux densities $\left<q_{\rm AGN}\right>=2.0$ with rms scatter
$\sigma_{\rm AGN}=0.5$. The larger scatter is expected for radio-loud AGN as these
are not expected to be correlated as strongly and the smaller average $q$
value is caused by the stronger flux densities of
radio-loud AGN.

A cutoff of $q=1.8$ is often used as a diagnostic
to distinguish between star-forming galaxies and radio-loud AGN
\citep[eg.][]{condon02}, and it appears that the spectroscopic
classification of radio sources used in this work
agrees well with this diagnostic.
Only 10 $(<1{\rm \,per\,cent})$ of the 
SF galaxies in the combined 6dFGS-NVSS-FSC 
sample have $q<1.8$, the value below
which a galaxy is 3 times more radio loud than the mean for
star-forming galaxies. 
These 10 galaxies may come from a class
of ``composite'' radio source, whose radio emission is a mixture of both
star-forming and radio-loud AGN activity or they may be candidate ``taffy''
galaxy pairs which recently suffered a direct collision \citep[eg.][]{condon93}. 
The SF spectral classification of these
objects is retained in further analysis.

\subsection{Far-infrared colours}

\begin{figure}
\centering
\includegraphics[width=\linewidth]{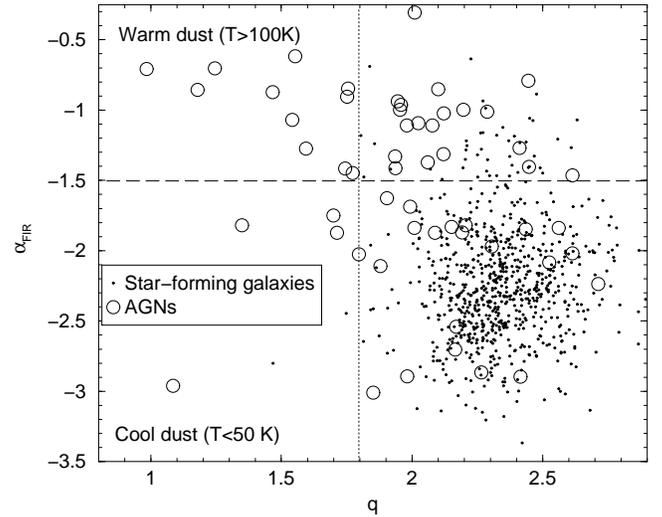}
\caption{ $\alpha_{\rm FIR}$ vs. $q$ (defined by equation~\ref{q})
  for the 769 star-forming galaxies
  (dots) and the 60 radio-loud AGN (circles) 
  in the combined 6dFGS-NVSS-FSC sample which also
  had a detection in the IRAS-FSC at 25\,$\mu$m. A dashed
  horizontal line is shown at
  $\alpha_{\rm FIR}=-1.5$ to delineate the regions occupied by ``warm''
  ($\alpha_{\rm FIR}>-1.5$) and ``cool'' ($\alpha_{\rm FIR}<-1.5$) 
  IRAS galaxies. The canonical value of $q=1.8$, used to distinguish
  star-forming galaxies from AGNs in other surveys \citep[eg.][]{condon02} 
  is shown as a dotted vertical line.}
\label{firsialpha}
\end{figure}

Another diagnostic which can be
used to distinguish between radio-loud AGN and
star-forming galaxies is the far-infrared spectral index between 60
and 25\,$\mu$m, $\alpha_{\rm IR} = \log(S_{25\,\mu{\rm m}}/S_{60\,\mu{\rm
    m}})/\log(60/25)$. $\alpha_{\rm IR}$ is a measure of the temperature of
the FIR emitting dust. Star-forming galaxies have
cooler ($T<50$\,K) dust emission because of their extended star and gas
distributions and therefore have values of $\alpha_{\rm IR}<-1.5$. 
The central engines of AGNs heat the dust to warmer
temperatures ($T>100$\,K) and have values of $\alpha_{\rm IR}>-1.5$ 
\citep{grijp92}. 

Sixty radio-loud AGN and 769 
SF galaxies in the sample were detected in the
IRAS\,FSC at 25\,$\mu$m. 
Fig.~\ref{firsialpha} plots $\alpha_{\rm IR}$ vs. $q$ for these
galaxies, comparing the dust temperature of each galaxy with its degree of
radio excess. The vast bulk of SF galaxies lie in the lower right
of the plot as expected from the radio-FIR correlation. The majority of
radio-loud AGN have $\alpha_{\rm IR}>-1.5$ (many lie outside 
the boundary on the upper left of Fig.~\ref{firsialpha}) 
indicating that the FIR spectral index, used
in concert with the radio-FIR correlation agrees well with spectral
classification as a discriminator of star-forming galaxies and radio-loud
AGN. 

\subsection{Composite optical spectra and IRAS comparisons}

For about 90\,per\,cent of the radio-detected 6dFGS galaxies in 
Table~\ref{datatable}, a 
classification as either AGN or SF (as discussed in Section~\ref{specclass}) appears 
straightforward.  Most of the galaxies we have classified as SF are 
detected by IRAS, and almost all of them lie in the lower 
right-hand quadrant of Fig.~\ref{firsialpha}, as expected if their FIR emission 
arises from dust heated in star-forming regions.  In contrast, the 
IRAS detection rate of the objects classified as Aa in Table~\ref{datatable} is 
close to zero, as expected if their inner regions are largely 
free of gas and dust and their radio emission comes exclusively from 
an AGN.

For the 10\,per\,cent of galaxies in Table~\ref{datatable} which show optical emission lines 
with ratios characteristic of an AGN, and which are therefore classified 
as Aae or Ae, the situation is less clearcut.  
As discussed by \citet{best05}, a possible problem with spectral
classification of emission line AGN as radio-loud is that emission-line 
AGN activity is often be accompanied by star formation \citep{kauffmann03}. 
Even if the AGN is radio-quiet, this associated star formation will
give rise to radio emission. For any such galaxy included in the sample,
the origin of the radio emission will be star formation, but the optical 
spectrum could still be dominated by emission lines from a (radio-quiet) 
AGN, leading us to classify the object as Aae or Ae. 

Although the present data are insufficient to disentangle
the separate contributions from star formation and the AGN to the radio
emission from these galaxies, we can estimate what fraction of them might have been
misclassified spectroscopically by comparing our spectroscopic classification
with the radio-FIR diagnostic.
High resolution Very Long Baseline Interferometry (VLBI) imaging of these sources
may be used to distinguish the parsec-scale AGN components of these galaxies
from the kpc scale radio structure resulting from star formation \citep[eg.][]{roy92}.
Such observations are outside the scope of the present paper.

Twenty spectroscopically classified AGN in the 6dFGS-NVSS sample lie in the
lower right region of Fig.~\ref{firsialpha} and would be classified as 
star-forming galaxies by the radio-FIR diagnostic. This implies that
about 1/3 of all spectroscopically classified
emission-line AGN (Ae and Aae class) in the 6dFGS-NVSS sample ($\sim$ 250) fail 
the radio-FIR diagnostic. It is probable that 
some of these may be genuinely misclassified
as radio-loud AGN from their optical spectra 
and derive the majority of their radio emission 
from star formation.
Assuming that all absorption-line radio-loud AGN (Aa class)
have been correctly classified, the disagreement between
spectroscopic classification and the radio-FIR diagnostic for 6dFGS-NVSS
AGN is $\sim 10$\,per\,cent. We estimated a similar reliability for our 
classification scheme in Section~\ref{specclass} from repeated visual 
inspection of the spectra. It is clear that difficulties in classifying 
emmision-line spectra with composite or borderline AGN+SF properties are the primary cause
of misclassification of radio sources in our sample.
We believe that this is a worst-case
estimate of the reliability of our spectroscopic classification 
and note that on the whole
the spectroscopic classification technique
agrees well with the radio-FIR correlation diagnostic.

Only 38\,per\,cent of
objects in the 6dFGS-NVSS sample were detected in the 6dFGS-FSC sample. This makes
the radio-FIR correlation diagnostic unsuitable for the sample presented in this
paper because a non-detection of
a galaxy in the 6dFGS-FSC sample does not preclude it from being a star-forming
galaxy in the 6dFGS-NVSS sample. Deeper wide-angle far-infrared
surveys are required to detect star-forming galaxies out to the redshifts
probed by the current generation of optical redshift surveys.

\subsection{The FIR-Radio correlation of 6dFGS galaxies}

\begin{figure}
\centering
\includegraphics[width=\linewidth]{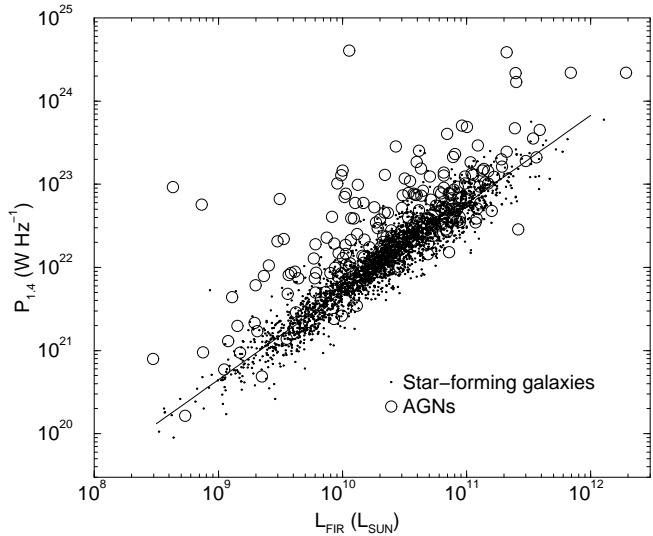}
\caption{
  The radio-FIR correlation of all 2\,690 star-forming galaxies (dots) and
  208 radio-loud AGN (open circles) in the combined 6dFGS-NVSS-FSC sample. A best fit line
  (equation~\ref{correqn} in the text) based on a subsample of 2\,242
  star-forming galaxies with accurate 60 and 100\,$\mu$m flux densities is
  also shown.}
\label{correlation}
\end{figure}

Fig.~\ref{correlation} shows the radio-FIR correlation
  ($P_{1.4}$ vs. $L_{\rm FIR}$) of 6dFGS-NVSS galaxies 
with a 60\,$\mu$m detection in the
IRAS\,FSC. Spectroscopically classified AGN are plotted as circles
and SF galaxies are plotted as dots. 
The SF galaxies show a strong correlation between radio power and FIR luminosity 
extending over 4 orders of magnitude though the scatter
  tends to increase above $L_{\rm FIR}=10^{11}\,L_{\odot}$ as was
  seen for IRAS-FSC detected galaxies in the  2dFGRS-NVSS sample of
  \citet{2dfnvss}. 
  The FIR luminosities show less correlation with radio power for
  radio-loud AGN; these are found across the whole range
  of FIR luminosity sampled. Many of them have ``excess'' radio powers,
  causing them to lie above the correlation defined by the star-forming
  galaxies. These radio-excess IRAS galaxies are of interest for
  understanding the connection between star formation and AGN in galaxies 
  \citep{drake03}. 

    A least-squares line of best fit to the subsample of 2\,242 star-forming
  galaxies for which accurate 60 and 100\,$\mu$m flux densities are
  available in the IRAS-FSC is also shown in Fig.~\ref{correlation}. The
  line has the form
\begin{equation}
\label{correqn}
\log(P_{1.4})=(1.06\pm0.01)\log(L_{\rm FIR}) + (11.11\pm0.1).
\end{equation}
which is similar to that found in other derivations of the radio-FIR
correlation for brighter galaxies (eg. \citet{condon91,cb98,de89}).

\subsection{Ultraluminous infrared galaxies}

  Ultra-luminous Infrared Galaxies (ULIRGs) with
  $L_{\rm FIR}>10^{12}\,L_{\odot}$ which are powered by a mixture of
  starburst activity and an AGN
  \citep{sanders96} are of interest for understanding the
  starburst-AGN connection. There are two ULIRGs in the present sample;
  such objects are extremely rare in the local universe. 
  The nearest known (Arp 220) has
  redshift $0.018$. The local ULIRG space density, estimated from the FIR
  luminosity function of \citet{saunders90}, is $<10^{-8}$\,Mpc$^{-3}$. The
  enclosed volume of the 6dFGS-NVSS sample (see Table~\ref{samplecompare})
  is 
  $\sim 3.9 \times 10^{8}$\,Mpc$^{3}$, meaning that less than 4 ULIRGs are
  expected
  in the 6dFGS-NVSS volume. Interestingly, \citet{2dfnvss} found 7
  ULIRGs in the 2dFGRS-NVSS sample, in a volume of $5.0\times10^7$\,Mpc$^3$
  in the redshift range $0.15<z<0.3$,
  which implies their space density is $1.4\times10^{-7}$\,Mpc$^{-3}$ within
  the more distant 2dFGRS redshift range. This suggests that these objects
  evolve strongly with redshift and is consistent with their measured evolution of
  $(1+z)^7$ between $0<z<1$ \citep{cowie04,sanders04}.

\section{The local Radio Luminosity Function}
\label{primlumfunc}

The luminosity function ($\Phi(P)$) of a sample of astronomical objects is a
measure of the variation in their
space density with luminosity. 
The \textit{local} radio luminosity function (RLF) 
is the global average space density of radio sources at the present
epoch \citep{auriemma77,condon89}. An accurate derivation of the local radio
luminosity function is important as a present epoch 
benchmark for studies of the population of radio sources at higher
redshift so that their cosmological evolution can be determined. 
In this section we derive the local radio luminosity
function (RLF) for the 6dFGS-NVSS
sample as a whole and for the star-forming galaxy and
radio-loud AGN subgroups separately. 

\subsection{Completeness}

The RLF can be calculated directly but 
requires a sample which is complete to the limits of all the surveys from
which it is derived. We therefore must define a 
subsample of the 6dFGS-NVSS sample which is complete to both its 
$K$-band magnitude limit and 1.4\,GHz flux density limit. 
In the $K$-band, the 2MASS\,XSC from which the 6dFGS primary sample
is selected is complete and reliable well below the $K=12.75$\,mag. limit of the 6dFGS 
\citep{2mass,jones06}.
The 6dFGS primary sample
may also miss populations of radio sources in the local 
universe either because their hosts appear stellar on optical plates 
(eg. QSOs, BL\,Lacs or compact galaxies) or because they are too blue in colour to be
detected in the 6dFGS primary sample (eg. Seyfert galaxies). 
A selection of such objects
were observed as additional targets during the 6dFGS under programme id. 125
\citep{jones04}. A full analysis of them will be the subject of a future paper.

The sharp drop in the source counts of NVSS radio sources below 2.8\,mJy 
(Fig.~\ref{primsc}) indicates that the NVSS catalogue is incomplete below this
limit. We applied a 1.4\,GHz flux density
limit of 2.8\,mJy to the 6dFGS-NVSS sample when calculating the RLF. The same 
flux density limit was 
used by \citet{2dfnvss} to measure the luminosity function of
2dFGRS-NVSS radio sources. There are 6\,961 6dFGS-NVSS
galaxies with $S_{1.4\,{\rm GHz}}\geq2.8$\,mJy, $K\leq12.75$ 
of which 6\,667 have 6dFGS\,DR2 measured redshift $z>0.003$.
Of these, 3\,997 were assigned spectral class SF, 2\,652 were assigned AGN and 18
were unclassified. 
These 18 unclassified spectra with measured
redshifts were classified on the basis of their measured radio power:
those with $P_{1.4}\leq10^{23}$\,W\,Hz$^{-1}$ (9) were classed as SF and
those with
$P_{1.4}>10^{23}$\,W\,Hz$^{-1}$ (9) were classed as AGN. 

We corrected the measured luminosity function for incomplete sampling
of the celestial sphere by the 6dFGS\,DR2 and for spectroscopic
incompleteness of the 6dFGS-NVSS sample.
To account for incomplete sampling of the celestial sphere we 
normalised all
volumes by the effective area of the 6dFGS-NVSS sample
derived in Section~\ref{primcov}. Of the 6\,961 
6dFGS-NVSS objects meeting our selection criteria, 280
had spectra which were too poor in quality to determine a
redshift and 14 were associated with galactic stars.
The poor quality spectra arose for a myriad of reasons which
were primarily instrumental (eg. broken fibres, misplaced buttons etc.) and
are expected to be a random subset of the data.
The spectroscopic incompleteness of 4\,per\,cent causes the luminosity
function to be underestimated, so values of $\Phi(P_{1.4})$ have been increased by 
4\,per\,cent to compensate. 

\subsection{Calculating the luminosity function}
\label{calcrlf}

We have measured the 
radio luminosity function using the $1/V_{\rm max}$ 
method of \citet{schmidt68}. $V_{\rm max}$ is the maximum volume in which
a galaxy will satisfy all of the sample selection criteria, which in the
case of the 6dFGS-NVSS sample are $S_{1.4\,{\rm GHz}}\geq2.8$\,mJy, $K\leq12.75$\,mag. and
$z>0.003$.

We have corrected the measured RLF for galaxy clustering
at a distance $s$ centred on our own Galaxy using
\begin{equation}
\label{lumfunccluster}
\frac{\rho_P}{\rho}=1+\frac{3}{3-\gamma_s}\left(\frac{s_0}{s}\right)^{\gamma_s}
\end{equation}
\citep{peebles80}, as discussed by \citet{condon02}.
Here $\rho_P / \rho$ is the expected overdensity near our own
Galaxy, or the space density ($\rho_P$) of \textit{local} galaxies divided by the
average space density ($\rho$) of \textit{all} galaxies, $\gamma_s$ is the
slope and $s_0$ the correlation scale length from a power-law
fit, $\xi\left(s\right)=\left(s/s_0\right)^{-\gamma_s}$, of the two-point
correlation function in redshift space. The two-point correlation function
of local radio sources has been derived from a subset of the 6dFGS-NVSS sample 
\citep{mauchthesis},
and is adequate to describe clustering of radio sources centred on our own
galaxy. Derivations of $\gamma_s$ and $s_0$ for radio sources in the
6dFGS-NVSS sample yielded
values of $\gamma_s=1.57$ and $s_0=10.07$. $\xi\left(s\right)$
has a power-law form for distances $s<30$\,Mpc, and
$\xi\left(s\right)\approx0$ for $s>30$\,Mpc.
Therefore to correct for the local
overdensity, we have multiplied the volume within $s$ by
equation~\ref{lumfunccluster} to calculate $V_{\rm max}$ for $s<30$\,Mpc. 
In practise, $V_{\rm max}$ is
always much larger than the local clustering volume, and so this
correction has made little difference to the results.

\subsection{Results}

\begin{table*}
\centering
\caption{Local RLFs at 1.4\,GHz for the radio-loud AGN, 
star-forming galaxies and for all radio sources.}
\label{lumfuncresults}
\begin{tabular}{lrrcrrcrrc}
\hline
 \multicolumn{1}{c}{} & \, & \multicolumn{2}{c}{All galaxies} & \, &
 \multicolumn{2}{c}{Star-forming galaxies} & \, & \multicolumn{2}{c}{Radio-loud AGN} \\
 \cline{3-4} \cline{6-7} \cline{9-10}
 \multicolumn{1}{c}{$\log_{10}\,P_{1.4}$} & & $N$ & \multicolumn{1}{c}{$\log\,\Phi$}
 & & $N$ & \multicolumn{1}{c}{$\log\,\Phi$} & & $N$ &
 \multicolumn{1}{c}{$\log\,\Phi$} \\
 \multicolumn{1}{c}{(W Hz$^{-1}$)} & & &
 \multicolumn{1}{c}{(mag$^{-1}$\,Mpc$^{-3}$)} & & &
 \multicolumn{1}{c}{(mag$^{-1}$\,Mpc$^{-3}$)} & & &
 \multicolumn{1}{c}{(mag$^{-1}$\,Mpc$^{-3}$)} \\
\hline
20.0 & & 3 & $-2.90^{+0.20}_{-0.39}$ & & 3 & $-2.90^{+0.20}_{-0.39}$ & & & \\
20.4 & & 46 & $-2.52^{+0.08}_{-0.10}$ & & 43 & $-2.56^{+0.09}_{-0.11}$ & & 3 & $-3.63^{+0.23}_{-0.54}$ \\
20.8 & & 116 & $-2.84^{+0.04}_{-0.05}$ & & 103 & $-2.89^{+0.04}_{-0.05}$ & & 13 & $-3.77^{+0.11}_{-0.15}$ \\
21.2 & & 319 & $-2.82^{+0.02}_{-0.03}$ & & 296 & $-2.85^{+0.03}_{-0.03}$ & & 23 & $-4.01^{+0.09}_{-0.11}$ \\
21.6 & & 654 & $-3.00^{+0.02}_{-0.02}$ & & 589 & $-3.05^{+0.02}_{-0.02}$ & & 65 & $-4.04^{+0.05}_{-0.06}$ \\
22.0 & & 1266 & $-3.25^{+0.01}_{-0.01}$ & & 1106 & $-3.31^{+0.01}_{-0.01}$ & & 160 & $-4.18^{+0.04}_{-0.04}$ \\
22.4 & & 1496 & $-3.68^{+0.01}_{-0.02}$ & & 1119 & $-3.79^{+0.02}_{-0.02}$ & & 377 & $-4.35^{+0.03}_{-0.03}$ \\
22.8 & & 1138 & $-4.23^{+0.02}_{-0.02}$ & & 588 & $-4.45^{+0.02}_{-0.02}$ & & 550 & $-4.62^{+0.02}_{-0.03}$ \\
23.2 & & 658 & $-4.78^{+0.02}_{-0.02}$ & & 133 & $-5.36^{+0.04}_{-0.05}$ & & 525 & $-4.91^{+0.03}_{-0.03}$ \\
23.6 & & 378 & $-5.06^{+0.03}_{-0.03}$ & & 26 & $-6.12^{+0.11}_{-0.14}$ & & 352 & $-5.09^{+0.03}_{-0.03}$ \\
24.0 & & 259 & $-5.36^{+0.03}_{-0.03}$ & & & & & 259 & $-5.36^{+0.03}_{-0.03}$ \\
24.4 & & 183 & $-5.57^{+0.04}_{-0.04}$ & & & & & 183 & $-5.57^{+0.04}_{-0.04}$ \\
24.8 & & 81 & $-6.03^{+0.06}_{-0.07}$ & & & & & 81 & $-6.03^{+0.06}_{-0.07}$ \\
25.2 & & 49 & $-6.33^{+0.08}_{-0.09}$ & & & & & 49 & $-6.33^{+0.08}_{-0.09}$ \\
25.6 & & 16 & $-6.74^{+0.16}_{-0.27}$ & & & & & 16 & $-6.74^{+0.16}_{-0.27}$ \\
26.0 & & 3 & $-7.30^{+0.25}_{-0.68}$ & & & & & 3 & $-7.30^{+0.25}_{-0.68}$ \\
26.4 & & 2 & $-8.12^{+0.27}_{-0.85}$ & & & & & 2 & $-8.12^{+0.27}_{-0.85}$ \\
 \cline{3-3} \cline{6-6} \cline{9-9}
Total & & 6667 & & & 4006 & & & 2661 & \\
$\left<V/V_{\rm max}\right>$ & & \multicolumn{2}{c|}{$0.518\pm0.004$} & &
\multicolumn{2}{c|}{$0.509\pm0.005$} & &
\multicolumn{2}{c}{$0.532\pm0.006$} \\
\hline
\end{tabular}
\end{table*}

\begin{figure}
\centering
\includegraphics[angle=-90,width=\linewidth]{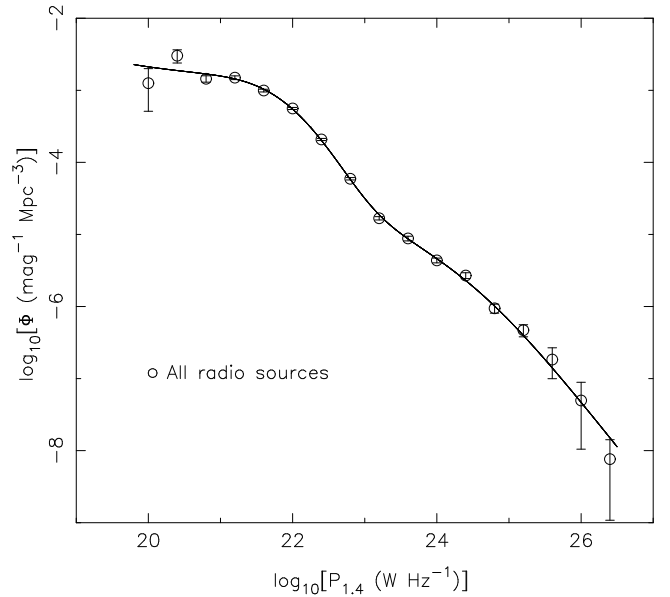}
\caption{The local luminosity function at 1.4\,GHz measured from 
	all the radio sources in
  the 6dFGS-NVSS sample. 
  The curve is the sum of the contributions to the luminosity
  function from fits of equations~\ref{lumfitsf}~and~\ref{lumfitagn} to the
  SF and AGN data respectively.}
\label{lumfuncplot}
\end{figure}

\begin{figure}
\centering
\includegraphics[angle=-90,width=\linewidth]{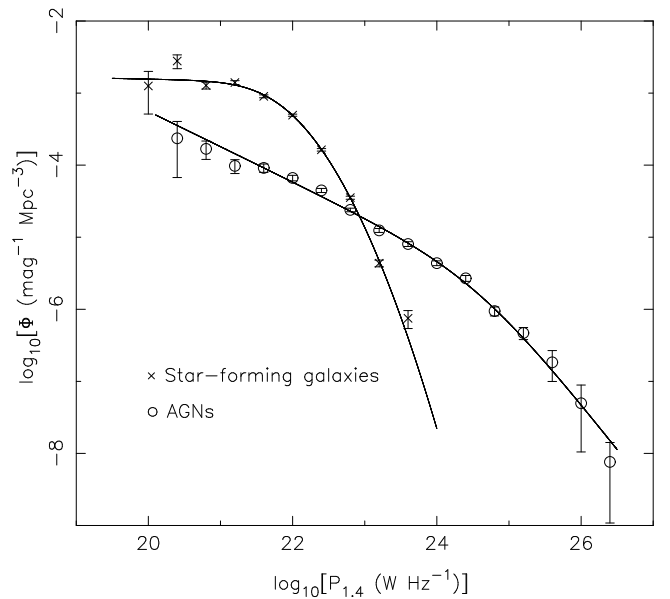}
\caption{The local luminosity function at 1.4\,GHz derived separately for the 
  radio-loud AGN
  (\textit{circles}) and star-forming galaxies (\textit{crosses}) in the
  6dFGS-NVSS sample. The two
  curves are the fits of equations~\ref{lumfitsf}~and~\ref{lumfitagn} to the 
  SF and AGN data respectively.}
\label{sfagnlumfunc}
\end{figure}

Table~\ref{lumfuncresults} lists the measured local radio luminosity
function for the star-forming galaxies, radio-loud AGN and the 6dFGS-NVSS sample as a
whole. The 4\,006 star-forming galaxies have $\left<V/V_{\rm
    max}\right>=0.509\pm0.005$, the 2\,661 radio-loud AGN have $\left<V/V_{\rm
    max}\right>=0.532\pm0.006$ and all the 6\,667 radio sources in the combined sample have
$\left<V/V_{\rm max}\right>=0.518\pm0.004$. The $\left<V/V_{\rm
    max}\right>$ value for radio-loud AGN is more than $3\sigma$ from the value of $0.5$ expected
if there were no significant clustering or evolution in the sample. Some
evolution of the radio-loud AGN population is probable over the 1--2\,Gyr lookback
time of the 6dFGS-NVSS sample.
The local radio
luminosity function of all 6dFGS-NVSS galaxies is shown in
Fig.~\ref{lumfuncplot}; its statistical errors are of order 1\,per\,cent or less
over 5 decades of radio luminosity. Separate local radio luminosity
functions for both star-forming galaxies and radio-loud AGN are shown in
Fig.~\ref{sfagnlumfunc}. These cross over at
$P_{1.4}=10^{23}$\,W\,Hz$^{-1}$, star-forming galaxies dominate the
population of radio sources below this power and radio-loud AGN dominate the
population above it.

Galaxy luminosity functions are commonly fitted by the Schechter function
\citep{schechter76}. This function turns over more steeply toward high
luminosities than radio luminosity functions of star-forming galaxies. 
Instead the radio luminosity
function is commonly fitted by the parametric form given by
\begin{equation}
\label{lumfitsf}
\Phi(P)=C \left(\frac{P}{P_{\star}}\right)^{1-\alpha}\exp
 \left\{-\frac{1}{2}\left[\frac{\log_{10}\left(1+P/P_{\star}\right)}{\sigma}\right]^2\right\}
\end{equation}
as was used to fit the luminosity function of IRAS galaxies by
\citet{saunders90}. The best-fitting parameters of equation~\ref{lumfitsf} for 6dFGS-NVSS
star-forming galaxies are: 
\begin{eqnarray}
C & = & 10^{-2.83 \pm 0.05}\,{\rm mag.^{-1}\,Mpc^{-3}}; \nonumber \\
P_{\star} & = & 10^{21.18 \pm 0.22}\,{\rm W\,Hz^{-1}}; \nonumber \\
\alpha & = & 1.02\pm0.15; \nonumber \\
\sigma & = & 0.60\pm0.04. \nonumber
\end{eqnarray}
Equation~\ref{lumfitsf} is inadequate to describe the RLF of 
radio-loud AGN and we have instead fitted these data with a 2 power-law function 
analogous to the optical luminosity function of quasars:
\begin{equation}
\label{lumfitagn}
\Phi(P)=\frac{C}{(P_{\star}/P)^\alpha + (P_{\star}/P)^\beta}
\end{equation}
\citep{brown01,dunlop90}. The best fitting parameters of equation~\ref{lumfitagn} for
6dFGS-NVSS radio-loud AGN are: 
\begin{eqnarray}
C & = & 10^{-5.50 \pm 0.25}\,{\rm mag.^{-1}\,Mpc^{-3}}; \nonumber \\
P_{\star} & = & 10^{24.59 \pm 0.30}\,{\rm W\,Hz^{-1}}; \nonumber \\ 
\alpha & = & 1.27\pm0.18; \nonumber \\
\beta & = & 0.49\pm0.04. \nonumber
\end{eqnarray}
The curves
plotted in Fig.~\ref{sfagnlumfunc} show the fits of 
equations~\ref{lumfitsf}~and~\ref{lumfitagn} to the
star-forming galaxies and radio-loud AGN respectively. 
The radio luminosity function of all 6dFGS-NVSS galaxies is
the sum of the two contributions from star-forming galaxies and 
radio-loud AGN,
as shown in Fig.~\ref{lumfuncplot}. The curve in this figure is
the sum of the two separate fits to the radio luminosity function for
the star-forming galaxies and radio-loud AGN. 

\subsection{Discussion}

The radio luminosity function of active galaxies
maintains a power-law form for about 5 orders of magnitude before turning
over above $P_{1.4}>10^{24.5}$\,W\,Hz$^{-1}$. This power-law
form of the radio-loud 
AGN luminosity function can only continue to fainter powers
if it does not 
exceed the space density of the luminous galaxies in which radio-loud AGN
preferentially reside. Taking the best fit
Schechter function to the B-band luminosity for early-type galaxies 
in the 2dFGRS from \citet{madgwick02}, and assuming that radio-loud AGN typically
live in the brightest early type galaxies (see Fig.~\ref{primlumdist},
\textit{top}), the space density of early-type galaxies with
($L_B>L^\star_B$) is $(9.3\pm0.3) \times 10^{-4}$\,Mpc$^{-3}$ (for
$H_0=70$\,km\,s$^{-1}$\,Mpc$^{-1}$). The space density of radio-loud 
AGN with $P_{1.4}>10^{20.2}$\,W\,Hz$^{-1}$ from the
6dFGS-NVSS luminosity function of Fig.~\ref{sfagnlumfunc} is $\approx
8\times 10^{-4}$\,Mpc$^{-3}$. If the power-law form
for radio-loud AGN continues to fainter radio powers
a maximum must be reached at $P_{1.4} \approx 10^{19.5}$\,W\,Hz$^{-1}$, 
just beyond the
limit of the present data. Though there is some evidence (eg. from
Fig.~\ref{primlumdist}) that radio-loud AGN of lower radio power may live in less
luminous host galaxies which have higher space density. 

In recent models, in which the radio-loud AGN luminosity function
is interpreted as the distribution of the time spent by an AGN at a given radio
power, the strong variability of radio-loud AGN is related to the heating of cooling
flows in galaxies and clusters \citep[eg.][]{nipoti05,bestagn}. From a fit to the
radio luminosity function of 2dFGRS-FIRST radio sources, 
\citet{nipoti05}
predict that the AGN luminosity function will turn over at
$P_{1.4}=10^{20.4\pm0.1}$\,W\,Hz$^{-1}$, just at the faint limit of the
6dFGS-NVSS luminosity function. Such models 
are poorly constrained by observational data at present and
require further constraints on the relationship between the mechanical
luminosity of heating of cooling flows by radio jets in 
AGN as well
as a more detailed understanding of the X-ray luminosity function of AGN.

\subsection{Comparison of recent RLF measurements}

\begin{figure}
\centering
\includegraphics[width=\linewidth]{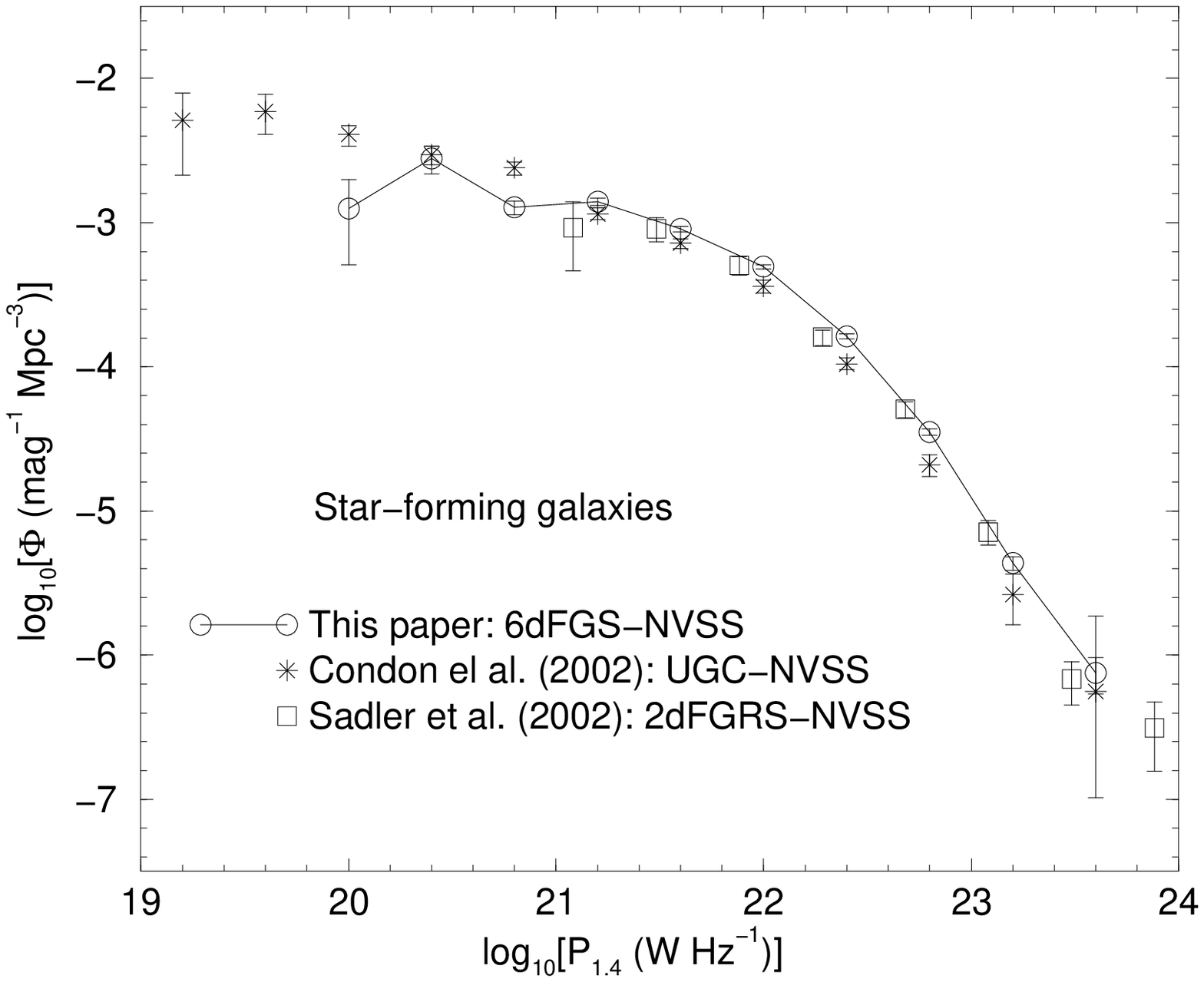}
\includegraphics[width=\linewidth]{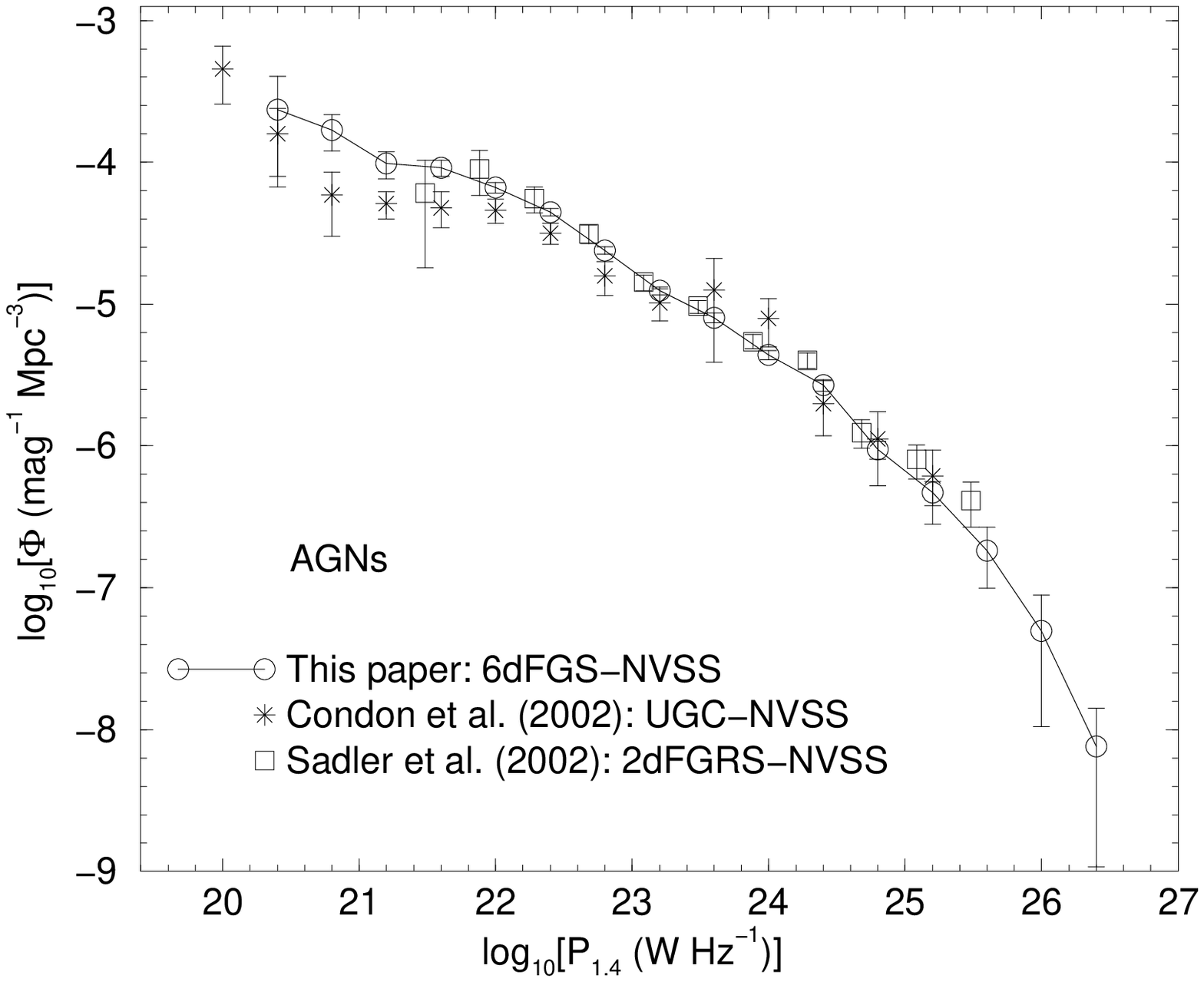}
\caption{The local radio luminosity function at 1.4\,GHz for star-forming galaxies
  (top) and radio-loud AGN (bottom), comparing results from three
  different samples.
  The UGC-NVSS luminosity functions of \citet{condon02}
  are shown as stars. The 2dFGRS-NVSS luminosity functions of
  \citet{2dfnvss} are shown as boxes and have been
  modified from $H_0=50$\,km\,s$^{-1}$\,Mpc$^{-1}$ to $H_0=70$\,km\,s$^{-1}$\,Mpc$^{-1}$ 
  by the scaling
  factor $P_{1.4} \propto H_0^{-2}$ and $\Phi \propto H_0^{3}$.
  The 6dFGS-NVSS luminosity functions are shown as
  circles with a line linking the points to highlight their position. }
\label{comparerlf}
\end{figure}

Fig.~\ref{comparerlf} shows a comparison of the RLFs
of 6dFGS-NVSS galaxies  
with those of UGC-NVSS galaxies and
2dFGRS-NVSS galaxies. Values of the 2dFGRS-NVSS luminosity function have been
shifted from their published values to $H_0=70$\,km\,s$^{-1}$\,Mpc$^{-1}$.
The properties of these three surveys are compared in
Table~\ref{samplecompare}. We make no comparison of our measured RLF with that
measured from
SDSS-NVSS/FIRST galaxies because the normalisation of the SDSS-NVSS/FIRST RLF 
is tied to the 2dFGRS-NVSS
RLF which it follows very closely over the complete range of radio powers
sampled \citep{best05}. We also do not make any comparison with the
2dFGRS-FIRST RLF because the FIRST survey is insensitive to the large-scale
radio emission seen in many nearby radio sources. We therefore
believe that \citet{maglio02}
have underestimated the total flux density of many of the
radio sources in the 2dFGRS-FIRST sample. 
The 6dFGS-NVSS sample presented here samples a larger
volume of space than the other radio source
samples and probes lookback times intermediate between the nearby
UGC-NVSS sample and the more distant 2dFGRS-NVSS one.
These three luminosity functions are all drawn from the
same radio source catalogue and differ only in the
optical/near-infrared samples from which they are selected. The agreement
between them is striking.

\subsubsection{Star-forming galaxies}

The 6dFGS-NVSS 
luminosity function of star-forming galaxies in the upper panel of
Fig.~\ref{comparerlf}
agrees with the
other two over the range $10^{20.5}<P_{\rm 1.4}<10^{22}$\,W\,Hz$^{-1}$.
The falloff at the
$P_{\rm 1.4}=10^{20}$\,W\,Hz$^{-1}$ 
point is caused by the small numbers of low luminosity star-forming galaxies 
in the 6dFGS
sample volume and can also be seen at the lowest radio
powers in the luminosity function of 2dFGRS-NVSS galaxies. The UGC-NVSS
luminosity function extends
about an order of magnitude deeper in radio power for star-forming
galaxies, reflecting the larger number of low-luminosity radio sources
detected in a flux density limited sample covering a larger solid angle. 
The 6dFGS-NVSS luminosity function of star-forming galaxies
does not fall as steeply
as the other two at the highest radio powers
($P_{1.4}>10^{22.5}$\,W\,Hz$^{-1}$). We believe this is caused by the
near-infrared selection of the 6dFGS-NVSS sample. 

\subsubsection{Radio-loud AGN}

The radio-loud AGN luminosity function of 6dFGS-NVSS galaxies agrees well
with the UGC-NVSS radio-loud AGN luminosity function. The UGC-NVSS radio-loud AGN sample
consists of only 294 objects and is therefore not as well constrained. Most of the
large error bars in the UGC-NVSS radio-loud AGN luminosity function overlap with the
6dFGS-NVSS points. A power-law fit
to the 6dFGS-NVSS luminosity function below $\log_{10}[P_{\rm 1.4}\,{\rm (W\,Hz^{-1})}]=25.0$ is
\begin{equation}
\log_{10}[\Phi(P_{\rm 1.4})]=(-0.61\pm0.02)\log[P_{\rm 1.4}] + (9.33\pm0.53)
\end{equation}
and to the 2dFGRS-NVSS luminosity function is
\begin{equation}
\log_{10}[\Phi(P_{\rm 1.4})]=(-0.62\pm0.03)\log[P_{\rm 1.4}] + (9.42\pm0.70).
\end{equation}
The 2dFGRS-NVSS luminosity function has the same power-law
form as the 6dFGS-NVSS luminosity function over nearly four orders of magnitude. 
The present data are insufficient to measure any possible evolution
over the narrow redshift range of $\tilde{z}=0.07$ 
to $\tilde{z}=0.14$ between the two samples.

\section{The star-formation density at the present epoch}

Radio emission from star-forming galaxies is the product of
massive ($M\geq8M_\odot$) stars which have short lifetimes
($\tau\sim10^7$\,yr). Roughly 90\,per\,cent of this radio emission
comes from non-thermal synchrotron electrons, which have been accelerated
in the remnants of type II supernovae, and the remainder comes from 
ionised hydrogen gas in H{\sc ii} regions heated by the most massive stars. This means that
the radio luminosity of star-forming galaxies is roughly proportional to the
rate of recent ($\tau \sim 10^8$\,yr) star-formation
\citep{condon92}. It is therefore possible to use the star-forming galaxies
in this sample to estimate the present epoch star-formation density of the
universe for galaxies with $K\leq12.75$\,mag., 
which is the zero point of the Madau diagram
\citep{madau96}. Estimates of the local star-formation density made from
radio surveys are particularly robust as they are free from the effects of
dust extinction. Effects of extinction are minimised in the 6dFGS-NVSS sample, 
as the radio selection is made from a near-infrared
galaxy sample. 

\subsection{Calculating the star-formation density} 

\begin{figure}
\centering
\includegraphics[width=\linewidth]{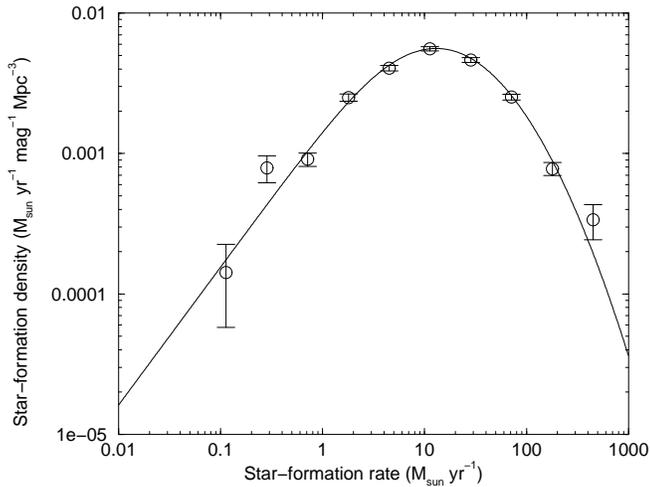}
\caption[The variation in star-formation density with star-formation rate
in the local universe.]
{\footnotesize
  The variation in star-formation density ($\rho_{\rm SF}({\rm SFR})$) 
  with star-formation rate for
  galaxies with star-formation rates between 0.1 and
  100\,$M_\odot$\,yr$^{-1}$ as derived
  from equation~\ref{sfdenseqn}. Values of $\Phi$ have been taken from
  Table~\ref{lumfuncresults} for star-forming galaxies with the errors
  scaled appropriately. The curve is derived from the fit of
  equation~\ref{lumfitsf} to the luminosity function of star-forming
  galaxies.}
\label{sfdensity}
\end{figure}

It is possible to estimate the local star-formation density as a function
of star-formation rate directly from the radio luminosity function using a
method described by \citet{cram98} and \citet{haarsma00}. Assuming a
typical Salpeter-like initial mass function of the form 
\begin{equation}
\label{salpeter}
\Psi(M)\propto M^{-2.35}
\end{equation}
between 0.1 and 100$M_\odot$, the star-formation rate of stars more massive
than 0.1$M_\odot$ (in $M_\odot$\,yr$^{-1}$) 
can be calculated from the 1.4\,GHz radio power by the relation 
\begin{equation}
\label{sfrateeqn}
{\rm SFR(P_{1.4})}=\frac{P_{1.4}}{8.85\times10^{20}}
\end{equation}
\citep{sullivan01}. The star-formation density can then be determined as a
function of star-formation rate by multiplying the star-formation rate
by the space density of radio sources, or simply
\begin{equation}
\label{sfdenseqn}
\rho_{SF}({\rm SFR})={\rm SFR}(P_{1.4})\times\Phi(P_{1.4})
\end{equation}
in $M_\odot$\,yr$^{-1}$\,mag.$^{-1}$\,Mpc$^{-3}$. Fig.~\ref{sfdensity}
shows equation~\ref{sfdenseqn} for the luminosity function of 6dFGS-NVSS
star-forming galaxies. Star-forming galaxies in the 6dFGS-NVSS sample trace
star-formation rates between 0.1 and 500\,$M_\odot$\,yr$^{-1}$, though the
major contribution to the local star-formation density comes from galaxies
with star-formation rates around 10\,$M_\odot$\,yr$^{-1}$ as has
been seen in other determinations of the star-formation density of the
local universe (eg. radio; \citet{2dfnvss} and H$\alpha$; \citet{gallego95}).

Integrating underneath the curve in Fig.~\ref{sfdensity} gives an
estimate of the global star-formation density at the present epoch
($\rho_{\rm SF}$). The
vast majority of radio sources in this sample contribute to the 
integral near its maximum at 10\,$M_\odot$\,yr$^{-1}$; the
less certain points below this maximum make little contribution to the integral.
This implies that the global star-formation density will be accurately
constrained by the data. Rather than compute the global star-formation
density from the binned data in Fig.~\ref{sfdensity} it is preferable to compute
$\rho_{\rm SF}$ directly from a sum over each sample galaxy. 
This is done by computing the local
power density function ($u(P_{1.4})$), 

The total radio power produced
per unit volume of space $U_{\rm SF}$ (in units of W\,Hz$^{-1}$\,Mpc$^{-3}$) 
is just the integral over all radio powers
of the power density function
\begin{equation}
U_{\rm SF}=\int_0^\infty u(P_{1.4})dP_{\rm 1.4},
\end{equation}
which can be calculated directly as the sum over all galaxies of
$P_{\rm 1.4}/V_{\rm max}$. $U_{\rm SF}$ is directly proportional to the
star-formation rate density $\rho_{\rm SF}$ and is computed using the
relation
\begin{equation}
\rho_{\rm SF}=1.13 \times 10^{-21} U_{\rm SF}
\end{equation}
the factor $1.13 \times 10^{-21}$ comes from the conversion between
star-formation rate and radio power in equation~\ref{sfrateeqn}. 

\subsection{Results}

The global power density of star formation at the present epoch implied by
the 6dFGS-NVSS sample is 
$U_{\rm SF}=(1.91\pm0.09)\times10^{19}$\,W\,Hz$^{-1}$\,Mpc$^{-3}$. 
This translates to a global star-formation density 
\begin{equation}
\label{globsfdensity}
\rho_{\rm SF}=\left(0.022\pm0.001\right)\hspace{0.5cm}M_\odot\,{\rm
  yr}^{-1}\,{\rm Mpc}^{-3}
\end{equation}
The uncertainty in $\rho_{\rm SF}$
quoted here is purely statistical and takes no account of the
errors arising from the conversion from $P_{1.4}$ to star-formation rate in
equation~\ref{sfrateeqn}. 
Errors in this conversion come primarily from
predicting the star-formation rate from the Type II SNe
rate, which contributes 90\,per\,cent of the radio power at 1.4\,GHz
\citep{sullivan01}, from models of the total radio spectral energy per Type II
supernova and from the assumption of
a Salpeter-like initial mass function in equation~\ref{salpeter}.

Values for $\rho_{\rm SF}$ measured
from the UGC-NVSS and 2dFGRS-NVSS samples are
shown in Table~\ref{samplecompare}.
The value of $\rho_{\rm SF}$ for 6dFGS-NVSS galaxies is intermediate
between values those for
the other samples. \citet{condon02} derived $\rho_{\rm
  SF}=0.018\pm0.001\,M_\odot$\,yr$^{-1}$\,Mpc$^{-3}$ for the UGC-NVSS
sample of 1672 star-forming galaxies and \citet{2dfnvss} derived $\rho_{\rm
  SF}=0.031\pm0.006\,M_\odot$\,yr$^{-1}$\,Mpc$^{-3}$ (converted to 
$H_0=70$\,km\,s$^{-1}$\,Mpc$^{-1}$) for 242 star-forming galaxies from the
2dFGRS-NVSS sample. 
Differences in the measured values of $\rho_{\rm SF}$ 
between the three radio samples are likely to be caused 
by their different optical/near-infrared magnitude limits. 
Fainter and hence less massive galaxies can make a significant contribution
to the local star formation density \citep{brinchmann04}. For example,
from Fig. 15 of \citet{brinchmann04} we estimate that the 
$K=12.75$ magnitude limit of the 6dFGS-NVSS means the sample misses at least
$\sim0.004\,M_\odot$\,yr$^{-1}$\,Mpc$^{-3}$ of the local star-formation
density measured from H$\alpha$. The 2dFGRS-NVSS sample on the other hand 
has an optical limit about 2 magnitudes fainter than the 6dFGS-NVSS sample 
and misses only a negligible amount of local star formation. Further differences
between the three samples are probably the result of evolution in the
star-formation density over the narrow redshift range.

Uncertainties in the relationship between monochromatic radio power 
and star-formation rate do not affect models describing 
the relationship between
star-formation rate and luminosities at shorter wavelengths. 
However, dust extinction within the H{\sc ii} regions where
star formation occurs contributes
increasingly to the model uncertainties at these shorter wavelengths. 
The value of $\rho_{\rm SF}$ measured here has been compared
with other recent determinations
measured from a broad range of wavelengths by \citet{hopkins06}. The interested
reader is referred to this paper for more information. 
The general agreement (usually to within a factor of 2)
between values of $\rho_{\rm SF}$ at different wavebands is reassuring.

\section{Bivariate radio-NIR luminosity functions}
\label{fraclumfunc}

The local RLF of radio-loud AGN
presented in section~\ref{primlumfunc} contains large numbers of galaxies in most
bins of radio power, which enables us to split it
into statistically significant subclasses.
In particular, the local radio luminosity functions can be
broken into bins of absolute $K$-band magnitude ($M_K$) to examine the
radio properties of galaxies of different near-infrared luminosities. The
near-infrared selection of the 6dFGS-NVSS sample is of particular
importance here, as the $K$-band galaxy luminosity is an
extinction-free estimator of the mass of the old stellar population in 
galaxies. The vast majority of radio-loud AGN reside in
early type galaxies whose absolute
$K$-band magnitude gives an indication of their bulge
mass which in turn is correlated with the black hole mass
\citep[eg.][]{marconi03}. It is therefore possible to examine how the
properties of radio-loud AGN may vary with black-hole mass.

One of the first bivariate radio-optical luminosity functions was
determined by \citet{auriemma77} from a sample of radio detections
of bright elliptical galaxies. They showed that the probability of
an elliptical galaxy to host a radio-loud AGN increases strongly as a 
function of optical luminosity. Further studies
using larger samples of elliptical galaxies were carried out by \citet{sadler89}
and \citet{ledowen} who verified the result of \citet{auriemma77} and
found that the break in radio luminosity function scaled strongly with
host galaxy luminosity. More recently \citet{best06} used the SDSS-NVSS/FIRST
sample of 2\,215 galaxies to calculate accurate 
bivariate radio luminosity-mass functions, from 
both host-galaxy mass and black-hole mass. They found that the \textit{fraction} of
galaxies hosting a radio-loud AGN increases strongly as a function of both
galaxy and black hole masses. \citet{best06} also argued that there was no evidence for
any dependence of the break luminosity on black hole mass. 

Here, we measure the bivariate radio-NIR luminosity function for the first time
which we can compare with bivariate radio-optical luminosity functions. The
6dFGS-NVSS bivariate luminosity function is calculated from accurate 
\textit{observed} quantities measured from homogeneous all-sky surveys. We
therefore avoid any errors which may arise from calculating the bivariate luminosity
function from \textit{derived} quantities. For example the large observed scatter in the
$M_{\rm bh}-\sigma$ relation of 0.3\,dex \citep{mbhsigma} is a significant
source of error in the bivariate radio luminosity versus black-hole mass function 
measured by \citet{best06}.

\subsection{Calculating the bivariate luminosity function}
\label{bivcalculate}

When separating the radio luminosity
function into $M_K$ bins it is preferable to define a
normalised, or fractional luminosity function
\begin{equation}
\label{fractionallumfunc}
F_K(P_{1.4})=\frac{\phi_K(P_{1.4})}{\rho_K}
\end{equation}
which is the radio luminosity function of galaxies in $M_K$
bin $M_K\pm\Delta M_K$ ($\phi_K(P_{1.4})$ in Mpc$^{-3}$) divided by the volume
density ($\rho_K$ in Mpc$^{-3}$) of
\textit{all} 
galaxies with absolute magnitude in the $M_K$-magnitude bin
$M_K\pm\Delta M_K$. 
The radio luminosity function in
each $K$-band magnitude bin $\phi_K(P_{1.4})$ 
is calculated using the $1/V_{\rm max}$ method
\citep{schmidt68} 
in 1-magnitude radio luminosity bins of width 0.4 
in $\log(P_{1.4})$. This gives
$\phi_K(P_{1.4})$ in units of mag$^{-1}$\,Mpc$^{-3}$. 
With this definition $F_K(P_{1.4})dP_{1.4}$
represents the fraction of all galaxies in the absolute near-infrared
magnitude bin $M_K\pm \Delta M_K$ that have radio power between 
$P_{1.4}$ and $P_{1.4}+dP_{1.4}$.

To determine $\rho_K$, the volume density of galaxies with absolute
near-infrared magnitude $M_K$, it is convenient to use the $K$-band galaxy
luminosity function $\Phi_K$. It is important that the
$K$-band luminosity function used be subject to the same selection criteria
as the radio sample, so as not to introduce biases from cosmic variance 
into the data. 
We use the $1/V_{\rm max}$ determination of the 6dFGS K-band luminosity
function described by \citet{jones06}.
Its error bars are much smaller than those in the 6dFGS-NVSS luminosity
function and are therefore ignored when calculating $F_K(P_{1.4})$.

\subsection{The bivariate radio-NIR luminosity function of radio-loud AGN}
\label{agnbiv}

\begin{table*}
\centering
\caption{The bivariate radio-$K$-band luminosity function of radio-loud AGN.}
\label{agnbivtable}
\begin{tabular}{lrrcrrcrrcrrc}
\hline
  & \, & \multicolumn{2}{c}{$-23>M_K\geq-24$} & \, &
 \multicolumn{2}{c}{$-24>M_K\geq-25$} & \, & \multicolumn{2}{c}{$-25>M_K\geq-26$} 
 & \, & \multicolumn{2}{c}{$-26>M_K\geq-27$} \\
\multicolumn{1}{c}{$\log_{10}\,P_{1.4}$} & & \multicolumn{2}{c}{($\rho_K=1.88\times10^{-3}$)} & \, & \multicolumn{2}{c}{($\rho_K=1.14\times10^{-3}$)} &
\, &  \multicolumn{2}{c}{($\rho_K=2.10\times10^{-4}$)} & \, & \multicolumn{2}{c}{($\rho_K=9.92\times10^{-6}$)}\\
 \cline{3-4} \cline{6-7} \cline{9-10} \cline{12-13}
 (W\,Hz$^{-1}$) & & $N$ & \multicolumn{1}{c}{$\log_{10}[F_K(P_{1.4})]$}
 & & $N$ & \multicolumn{1}{c}{$\log_{10}[F_K(P_{1.4})]$} & & $N$ &
 \multicolumn{1}{c}{$\log_{10}[F_K(P_{1.4})]$} & & $N$ & \multicolumn{1}{c}{$\log_{10}[F_K(P_{1.4})]$} \\
\hline
20.4    & & 1 & $-1.49^{+0.30}_{-1.00}$   & & &        & & 1 & $-1.05^{+0.30}_{-1.00}$ & & & \\
20.8    & & 3 & $-1.70^{+0.20}_{-0.38}$   & & 7 & $-1.12^{+0.14}_{-0.21}$   & & 1 & $-1.34^{+0.30}_{-1.00}$ & & &\\
21.2    & & 7 & $-1.80^{+0.14}_{-0.22}$   & & 10 & $-1.45^{+0.12}_{-0.17}$  & & 4 & $-1.15^{+0.18}_{-0.31}$ & & &\\
21.6    & & 16 & $-1.93^{+0.10}_{-0.14}$  & & 38 & $-1.33^{+0.07}_{-0.08}$  & & 8 & $-1.28^{+0.14}_{-0.22}$ & & & \\
22.0    & & 31 & $-2.11^{+0.07}_{-0.09}$  & & 76 & $-1.62^{+0.05}_{-0.06}$  & & 49 & $-1.05^{+0.06}_{-0.07}$ & & & \\
22.4    & & 19 & $-2.59^{+0.09}_{-0.12}$  & & 189 & $-1.74^{+0.03}_{-0.03}$ & & 161 & $-1.11^{+0.04}_{-0.04}$ & & 5 & $-1.20^{+0.17}_{-0.29}$ \\
22.8    & & 8 & $-2.90^{+0.14}_{-0.20}$   & & 148 & $-2.08^{+0.04}_{-0.04}$ & & 380 & $-1.25^{+0.02}_{-0.02}$ & & 14 & $-1.49^{+0.11}_{-0.15}$ \\
23.2    & & 4 & $-3.35^{+0.18}_{-0.31}$   & & 57 & $-2.52^{+0.06}_{-0.07}$  & & 359 & $-1.46^{+0.02}_{-0.03}$ & & 105 & $-1.03^{+0.04}_{-0.05}$ \\
23.6    & & 2 & $-3.71^{+0.23}_{-0.54}$   & & 44 & $-2.61^{+0.07}_{-0.08}$  & & 209 & $-1.69^{+0.03}_{-0.03}$ & & 93 & $-1.22^{+0.04}_{-0.05}$ \\
24.0    & & &         & & 14 & $-3.21^{+0.10}_{-0.14}$  & & 167 & $-1.82^{+0.03}_{-0.04}$ & & 77 & $-1.31^{+0.05}_{-0.06}$ \\
24.4    & & &        & & 6 & $-3.51^{+0.15}_{-0.23}$   & & 99 & $-2.06^{+0.04}_{-0.05}$  & & 74 & $-1.31^{+0.05}_{-0.06}$ \\
24.8    & & &        & & 3 & $-3.89^{+0.20}_{-0.38}$   & & 33 & $-2.60^{+0.07}_{-0.09}$  & & 42 & $-1.57^{+0.07}_{-0.08}$ \\
25.2    & & &        & & 1 & $-4.42^{+0.30}_{-1.00}$   & & 10 & $-3.00^{+0.12}_{-0.17}$  & & 36 & $-1.67^{+0.07}_{-0.08}$ \\
25.6    & & &        & & 1 & $-4.17^{+0.30}_{-1.00}$   & & 4 & $-3.55^{+0.18}_{-0.31}$  & & 10 & $-2.33^{+0.12}_{-0.18}$ \\
26.0    & & & & & &                & & 2 & $-3.64^{+0.26}_{-0.73}$ & & & \\
 \cline{3-3} \cline{6-6} \cline{9-9} \cline{12-12}
Total & & 91 & & & 594 & & & 1487 & & & 456\\
\hline
\end{tabular}
\end{table*}

\begin{figure}
\centering
\includegraphics[width=\linewidth]{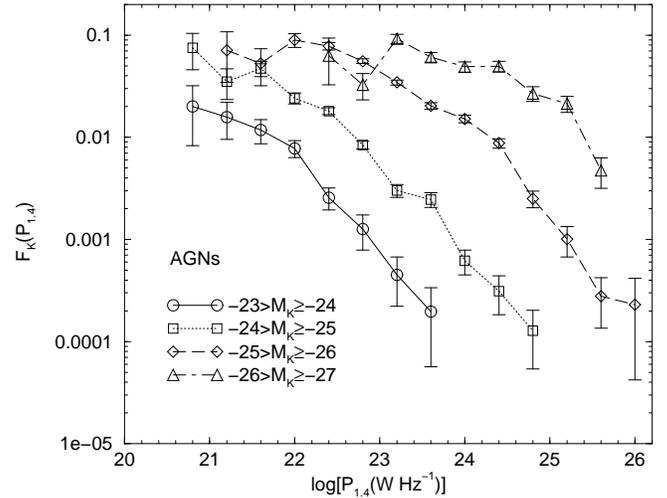}
\caption{The fractional luminosity function of radio-loud AGN calculated
  using the procedure described in section \ref{bivcalculate} of the text.
  $F_K(P_{1.4})$ is calculated in 4 bins of absolute $K$ magnitude ($M_K$) 
  as labelled in the figure. The two $M_K$ bins centred at $-25.5$ and $-26.5$ are
  brighter than $M_K^\star=-24.6$.}
\label{bivagnlumfunc}
\end{figure}

Table~\ref{agnbivtable} lists and Fig.~\ref{bivagnlumfunc} 
shows $F_K(P_{1.4})$ for radio-loud AGN in four
bins of $M_K$.
Below $F_K(P_{1.4})=0.1$ the broad break between $P_{1.4}=10^{24}$ and $10^{25}$\,W\,Hz$^{-1}$
seen in the RLF in Fig.~\ref{sfagnlumfunc} can be seen in the brightest $M_K$ bins ($M_K=-25.5,-26.5$).
In the faintest 2 bins ($M_K=-23.5,-24.5$), the data do not extend to 
high enough radio powers
to detect a break. In the brightest bins of $M_K$ the fractional luminosity 
function turns over below the radio power at which $F_K(P_{1.4})=0.1$. 
It is probable that this turnover is caused by
the normalisation of the fractional luminosity function; 
a larger fraction of radio sources cannot be measured at a given $M_K$ 
because nearly all available galaxies are detected. 
The turnover occurs at fainter radio powers
in less luminous galaxies, this is probably because both
less luminous galaxies and less powerful radio sources have
higher space density.

There is a clear decrease with radio power in the fraction
of galaxies which are radio-loud AGN for all $M_K$, reflecting the
decrease in space density of radio-loud AGN with increasing radio power. 
For a given radio power, there is a strong increase in the
fraction of galaxies which are radio-loud AGN with brightening $M_K$. 
More luminous galaxies are therefore more likely to host a
radio-loud AGN as has been seen in previous determinations of the
fractional radio-optical luminosity function of elliptical galaxies. 
To properly quantify this effect
we determine the fractional luminosity function in its integral form. 

\begin{figure}
\centering
\includegraphics[width=\linewidth]{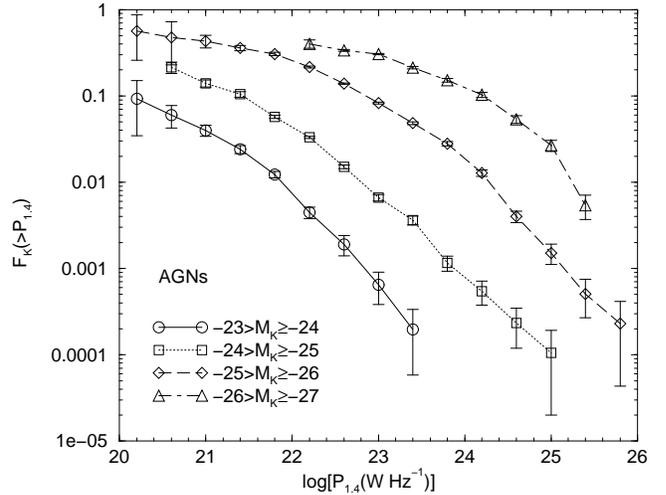}
\caption{The fractional luminosity function of radio-loud
  AGN in integral form. $F_K(>P_{1.4})$
  denotes the fraction of all galaxies which host a radio-loud AGN with 
  radio power greater than $P_{1.4}$. 
  Errors have been
  calculated using the bootstrap method with $10^6$ iterations.}
\label{intagnbivrlf}
\end{figure}

\subsection{The integral bivariate radio-NIR luminosity function of radio-loud AGN}

The integral form of the fractional luminosity function at a particular radio
power ($F_K(>P_{1.4})$) is calculated in each $M_K$ bin 
from the sum of $F_K(P_{1.4})$ between the largest bin of radio power
($P_{1.4}^{\rm max}$) to $P_{1.4}$.
It measures the
fraction of all galaxies which are radio sources with powers greater than
$P_{1.4}$ in each $M_K$ bin. It is not straightforward to calculate
the error at each point of $F_K(>P_{1.4})$ because the points are not
independent. For this reason we calculated errors in each radio power bin of
$F_K(>P_{1.4})$ using bootstrap analysis with $10^6$ iterations. 

Fig.~\ref{intagnbivrlf} shows the fractional luminosity function of 
radio-loud AGN in
integral form in four bins of $M_K$, with bootstrap errors. With
the available data, radio powers at which 40-50\,per\,cent of galaxies are detected
as radio-loud AGN are reached at the brightest $M_K$ bins. The function has
approximately the same shape for all bins of $M_K$
though this breaks down above $F_K(>P_{1.4})=0.3$ which 
is caused by the turnover in $F_K(P_{1.4})$ discussed in Section~\ref{agnbiv}. \citet{best06}
also found a ceiling in the bivariate radio luminosity-mass function of SDSS-NVSS/FIRST
galaxies at the 30\,per\,cent level.
We determine the dependence of $f_{\rm radio-loud}$ (the fraction of all galaxies
that are radio loud) on $L_K$ by scaling $F_K(>P_{1.4})$ to remove the dependence on
$M_K$. The best-fitting scaling factor for $F_K(>P_{1.4})<0.3$ is 
$f_{\rm radio loud} \propto L_K^{2.1}$; clearly, the probability of a galaxy to host radio-loud
AGN is a strong function of its $K$-band luminosity.

Given that the majority ($>70$\,per\,cent) of the radio-loud AGN in the 6dFGS-NVSS
sample reside in early-type galaxies, their total $K$-band luminosity should be
close to or equal to their near-infrared bulge luminosity ($L_{K,{\rm bul}}$). 
Assuming this,
the $K$-band luminosities of these radio-loud AGN are indicative of the
black-hole mass of the galaxy via the $M_{\rm BH}$-$L_{K,{\rm bul}}$ correlation 
\citep{marconi03}. The precise scaling is uncertain though, as it is unclear
at this stage of the precise contribution from disks to the $K$-band luminosity
of the radio-loud AGN in our sample.
An accurate
derivation of the $K$-band bulge luminosity of all 6dFGS galaxies is beyond the scope of this paper. 
Accurate velocity dispersions will be measured from spectra 
of the 15\,000
brightest elliptical galaxies in the 6dFGS\citep{jones04}. We expect
to detect about 30\,per\,cent of these in the NVSS (see Fig.~\ref{intagnbivrlf}). A full analysis
of these galaxies will be the subject of a future paper.

\section{Conclusion}
\label{primconclusion}

In this paper we have derived a catalogue of 7\,824 radio sources 
from the NVSS catalogue which have been observed in the 6dFGS\,DR2
$K$-selected primary sample. The main results of this paper are:

\begin{itemize}

\item  
  The median redshift of all radio sources in the sample is $\tilde{z}=0.046$, which is
  similar to that of the 47\,317 $K$-band selected galaxies from
  which the radio sources were selected.  
  60\,per\,cent of 6dFGS-NVSS primary targets are spectroscopically identified as
  star-forming galaxies which form a
  much more nearby population ($\tilde{z}=0.035$) and 40\,per\,cent of 6dFGS-NVSS
  primary targets are identified as radio-loud AGN which form a
  more distant population ($\tilde{z}=0.073$).

\item
  Star-forming galaxies and radio-loud AGN have quite distinct distributions in the
  plane of radio power vs. absolute $K$ magnitude. Star-forming galaxies
  tend to have radio powers $P_{1.4}<10^{23}$\,W\,Hz$^{-1}$, though they have
  a wide range of absolute $K$ magnitudes. The radio-loud AGN span a wide
  range in radio power but are almost all found in the
  most luminous near-infrared galaxies with $M_K>M_K^\star$.

\item
  2\,690 star-forming galaxies and 208 radio-loud AGN were detected at 60\,$\mu$m in
  the IRAS Faint Source Catalogue. 
  All of the star-forming galaxies with IRAS-FSC detections in the
  6dFGS-NVSS sample are found to lie on the radio-FIR correlation, with
  average FIR-radio flux ratio parameter $\left<q_{\rm SF}\right>=2.3$ with
  rms scatter $\sigma_{\rm SF}=0.18$, in close agreement with the results of
  other recent radio-FIR surveys \citep[UGC-NVSS;][]{condon02}. Many of the
  208 radio-loud AGN with IRAS-FSC 60\,$\mu m$ detections also lie on the radio-FIR
  correlation. However, these tend to have hotter IRAS 60-25\,$\mu$m
  colours. The combination of the radio-FIR correlation and the
  60-25\,$\mu$m 
  colour as a means of AGN/Star-forming classification for radio sources
  agrees with spectroscopic classification for more than $90$\,per\,cent of all galaxies,
  which is reassuring considering the differences in these classification
  techniques.

\item 
  We have measured accurate and homogeneous local radio luminosity functions at
  1.4\,GHz of star-forming galaxies and radio-loud AGN. The luminosity functions
  presented here agree well with recent determinations of the local radio
  luminosity function derived from different optical samples (eg. UGC-NVSS;
  \citet{condon02}, 2dFGRS-NVSS; \citet{2dfnvss}). 

\item
  The star-formation density at the present epoch has been measured from
  the 6dFGS-NVSS sample and was found to be
  $\rho_{SF}=(0.022\pm0.001)$\,M$_\odot$\,yr$^{-1}$\,Mpc$^{-3}$, in agreement
  with previously derived values at radio and other wavelengths.
  
\item
  We have split the radio luminosity function of radio-loud AGN
  into four bins of $M_K$ and compared with the
  $K$-band luminosity function of all 6dFGS galaxies to
  calculate a fractional luminosity function $F_K(P_{1.4})$.
  We find that $F_K(P_{1.4})$ for radio-loud AGN
  increases strongly as a function of near-infrared luminosity,
  which in turn is indicative of an increasing probability for a galaxy to
  host a radio-loud AGN with increasing black-hole mass.

\end{itemize}

\section*{Acknowledgements}

We would like to thank the Anglo-Australian Observatory
staff at the UK Schmidt Telescope and the entire 6dFGS team
for ensuring the success of the 6dFGS. We thank 
D. Heath Jones for providing early versions of the 
6dFGS $K$-band galaxy luminosity function for analysis. 
We thank the referee for carefully reading the paper and 
useful comments which improved the final version.
This research uses the NVSS radio survey, carried out using the 
National Radio Astronomy Observatory (NRAO) Very Large Array. The NRAO
is a facility of the National Science Foundation operated under cooperative
agreement by Associated Universities, Inc. Tom Mauch acknowledges the
financial support of an Australian Postgraduate Award. Elaine M. Sadler
acknowledges support from the Australian Research Council through
the award of an ARC Australian Professorial Fellowship.


\label{lastpage}
\end{document}